\documentclass[twocolumn]{aastex63}

\usepackage{amsmath}
\usepackage{multirow}
\usepackage{textcomp}
\usepackage{xfrac}
\usepackage{hyperref}

\newcommand{\Rs}{R$_\odot$}

\newcommand{\mic}{\textmu m}

\newcommand{\term} [4] {$ {\rm {^#1}{#2^{#3}_{#4}}}$}
\newcommand{\rsun}{\mbox{\,$\rm R_{\odot}$}}        
\newcommand{\beq}{\begin{equation}}
\newcommand{\eeq}{\end{equation}}

\DeclareMathAlphabet{\mathsc}{OT1}{cmr}{m}{sc}
\def\testbx{bx}%
\DeclareRobustCommand{\mion}[2]{%
\relax\ifmmode
\ifx\testbx\f@series
{\mathbf{#1\,\mathsc{#2}}}\else
{\mathrm{#1\,\mathsc{#2}}}\fi
\else\textup{#1\,{\mdseries\textsc{#2}}}%
\fi}

\submitjournal{\textit{The Astrophysical Journal}}

\shorttitle{Multi-Wavelength Characterization of Coronal Plasma}
\shortauthors{Del Zanna et al.}

\begin{document}

\title{Coronal Densities, Temperatures and Abundances During the 2019 Total Solar Eclipse: \\The Role of Multi-Wavelength Observations in Coronal Plasma Characterization}
\correspondingauthor{Giulio Del Zanna}
\email{}

\author[0000-0002-4125-0204]{Giulio Del Zanna}
\affiliation{DAMTP, CMS, University of Cambridge, Wilberforce Road, Cambridge CB3 0WA, UK}

\author[0000-0002-4498-8706]{Jenna Samra}
\affiliation{Center for Astrophysics $|$ Harvard \& Smithsonian, 60 Garden Street, Cambridge, MA, 02138, USA}

\author{Austin Monaghan}
\affiliation{HAO, National Center for Atmospheric Research, P.O. Box 3000, Boulder CO 80307-3000, USA}

\author[0000-0001-8775-913X]{Chad Madsen}
\affiliation{Center for Astrophysics $|$ Harvard \& Smithsonian, 60 Garden Street, Cambridge, MA, 02138, USA}

\author[0000-0001-5681-9689]{Paul Bryans}
\affiliation{HAO, National Center for Atmospheric Research, P.O. Box 3000, Boulder CO 80307-3000, USA}

\author[0000-0001-7416-2895]{Edward DeLuca}
\affiliation{Center for Astrophysics $|$ Harvard \& Smithsonian, 60 Garden Street, Cambridge, MA, 02138, USA}

\author{Helen Mason}
\affiliation{DAMTP, CMS, University of Cambridge, Wilberforce Road, Cambridge CB3 0WA, UK}

\author{Ben Berkey}
\affiliation{HAO, National Center for Atmospheric Research, P.O. Box 3000, Boulder CO 80307-3000, USA}

\author[0000-0002-5084-4661]{Alfred de Wijn}
\affiliation{HAO, National Center for Atmospheric Research, P.O. Box 3000, Boulder CO 80307-3000, USA}

\author[0000-0002-8748-2123]{Yeimy J. Rivera}
\affiliation{Center for Astrophysics $|$ Harvard \& Smithsonian, 60 Garden Street, Cambridge, MA, 02138, USA}

\begin{abstract}
  The Airborne Infrared Spectrometer (AIR-Spec) offers an unprecedented opportunity to explore the
Near Infra-Red (NIR) wavelength range. It has been flown at two total solar eclipses, in 2017 and 2019. 
The wavelength range of the
much improved instrument on the second flight (July 2, 2019) was shifted to cover two density
sensitive lines from S XI. 
In this paper we study detailed
diagnostics for temperature, electron density and elemental abundances by comparing results from
AIR-Spec slit positions above the east and the west limb 
with those from Hinode/EIS, the PolarCam detector and SDO/AIA. 
We find very good agreement in the electron densities obtained from the EIS EUV line ratios,
those from the NIR S XI ratio and those obtained from the polarized brightness PolarCam 
measurements. Electron
densities ranged from Log N{\rm e} [cm$^{-3}]$ = 8.4 near the limb, falling to 7.2 at $R_0=1.3$.
EIS spectra indicate that the temperature distribution above the west limb
is near-isothermal at around 1.3 MK, while that on the east has an additional higher-T component.
The AIR-Spec radiances in Si X and S XI as well as the AIA data in the 171, 193, and 211~\AA\ 
bands are consistent with the EIS results. EIS and AIR-Spec data indicate 
that the sulphur abundance (relative to silicon)
is photospheric in both regions, confirming our previous results of the 2017
eclipse. The AIA data also indicate that the absolute iron abundance is photospheric. 
 Our analysis confirms the importance of
the diagnostic potential of the NIR wavelength range, and that this important wavelength range can
be used reliably and independently to determine coronal plasma parameters.
\end{abstract}


\keywords{Solar coronal lines (2038); Solar instruments (1499); Total eclipses
(1704); Spectrometers (1554)}

\section{Introduction} 
\label{sec:introduction}

The visible forbidden lines provided the first detailed information about the
plasma conditions in the solar corona, with several spectroscopic observations
carried out in the 1950's to 1970's \citep[see the review in][]{delzanna_deluca:2018}. 
Since then, most observations have been carried out at shorter wavelengths in the XUV.
However, there is now renewed interest in the forbidden lines in the visible and
especially in the near-infrared (NIR).  This is partly because these lines 
are significantly enhanced due to photo-excitation (PE) from the solar disk radiation,
hence  are easier to measure out to greater distances from the solar limb, compared to
the XUV lines, as shown e.g. by \cite{Habbal2011} using observations of visible lines.

There have been many studies from the ground based on the visible lines
during total eclipses, as for example \cite{ding_habbal:2017,koutchmy_etal:2019}.
For a recent review of such observations see \cite{habbal_etal:2021}.

In this paper, we focus on the NIR lines, as they
offer the possibility for routine
coronal magnetic field measurements via the  Zeeman and Hanle effects
\citep[see, e.g.][]{Judge1998,Judge2001}, partly thanks to recent improvements in NIR detectors.
Also, measurements of non-thermal widths are  easier to obtain in the NIR than the EUV
 (for a recent attempt to measure non-thermal line widths
in the EUV see \citealt{delzanna_etal:2019_ntw}). 
The NIR lines  offer in principle the same spectral diagnostic applications as the XUV lines, 
in terms of measurements of electron densities,  temperatures,
and  elemental abundances from line ratios \citep[see ][]{delzanna_deluca:2018}.
Such measurements are fundamental constraints to physical models of the middle corona,
  the almost unexplored region between 1.5 and 3 solar radii,  as reviewed by \cite{west_etal:2022}.

Densities and temperatures are obvious key parameters.
It has been recognised in the literature that it is also important to measure
elemental abundances. 
It is now well established that
coronal abundances, depending on the feature observed, are different
to the photospheric ones. Variations are strongly correlated with the
FIP (First Ionization Potential) of an element, with generally low FIP elements being enhanced, relative to
the high-FIP ones (the so-called FIP effect). The variations are likely related to the
plasma heating processes in the chromosphere
(where neutrals are ionized). Also, in principle they are not significantly modified once established
in the corona, so could be used as a tracer of the origin of the solar wind streams,
whenever  elemental abundances are measured in-situ. 
For details, see the Solar Living Reviews by
\cite{laming:2015} and \cite{delzanna_mason:2018}.

For the above (and other) reasons, several missions have recently included diagnostics of
NIR lines. Various space-based missions such as ESA's Proba-3 and the first Indian large solar
mission Aditya will observe several forbidden lines.
On the ground, UComp, an improved version of the successful Coronal Multi-Channel Polarimeter
(CoMP) is now operational \citep{Landi2016,Tomczyk2019,ucomp}. 
A major new facility is the Daniel K. Inouye Solar Telescope
(DKIST, \citealp{Tritschler2016,Rimmele2020}).  The main coronal instrument, the
Cryogenic Near Infrared Spectropolarimeter (Cryo-NIRSP, \url{https://atst.nso.edu/inst/cryonirsp})
will make regular observations, but only close to the solar limb and with a small field of view (FOV).
In addition, we note that any ground-based telescope is strongly 
limited by atmospheric absorption, except in a few spectral regions.

Cryo-NIRSP is planning to observe a few lines that are known to be observable,
but the NIR is in principle rich in other diagnostics. As the NIR is  an almost
unexplored spectral region in solar physics, the
Airborne Infrared Spectrometer (AIR-Spec, \citealt{Samra2021}) was funded by the NSF and
developed by Smithsonian Astrophysical Observatory (SAO) to explore the NIR. 

The instrument has observed the quiescent
solar corona during two eclipses, in 2017 and 2019, flying
on-board the NSF/NCAR Gulfstream V (GV) at an altitude of 12 km, where
atmospheric absorption is much reduced.  The long slit of the instrument allows
measurements of the outer corona, at far greater heights than Cryo-NIRSP.

The 2.853~\mic\ Fe IX line was detected for the first time during the
first flight \citep{Samra2018,Samra2019}. Multi-wavelength observations during the 
2017 eclipse
were presented in \cite{madsen_etal:2019}, and used to measure 
electron densities, ionization temperatures (i.e. temperatures obtained 
from lines of different ions, assuming ionization equilibrium), and  elemental abundances.
The results were  close to those obtained from an analysis
of earlier  SoHO SUMER and UVCS observations of  quiescent streamers during solar minimum
\citep{delzanna_etal:2018_cosie}.
The electron densities were obtained from EUV line ratios 
observed by the  Hinode Extreme Ultraviolet Imaging Spectrometer \citep[EIS,][]{culhane_etal:2007}.
Emission measures obtained from both AIR-Spec and EIS indicated a near-isothermal
temperature distribution, and, perhaps surprisingly, nearly photospheric abundances
\citep{asplund_etal:2009}
in terms of relative abundance between Sulphur and Silicon.

The second flight observed the July 2, 2019 total solar eclipse with a much
improved and modified instrument. The wavelength coverage was shifted so as
to observe the 1.4 \mic\ \ion{S}{11} line, to obtain electron densities
directly from the ratio with the 1.9 \mic\ \ion{S}{11} line.
The observations and preliminary results from this eclipse
are described in \cite{samra_etal:2022}. 
The observations were successful and the 1.4 \mic\ \ion{S}{11} line
was recorded for the first time, although careful atmospheric modeling
was required to measure its intensity.

\begin{figure*}[!htbp]
	\centering
	\includegraphics[angle=0,width=0.55\linewidth]{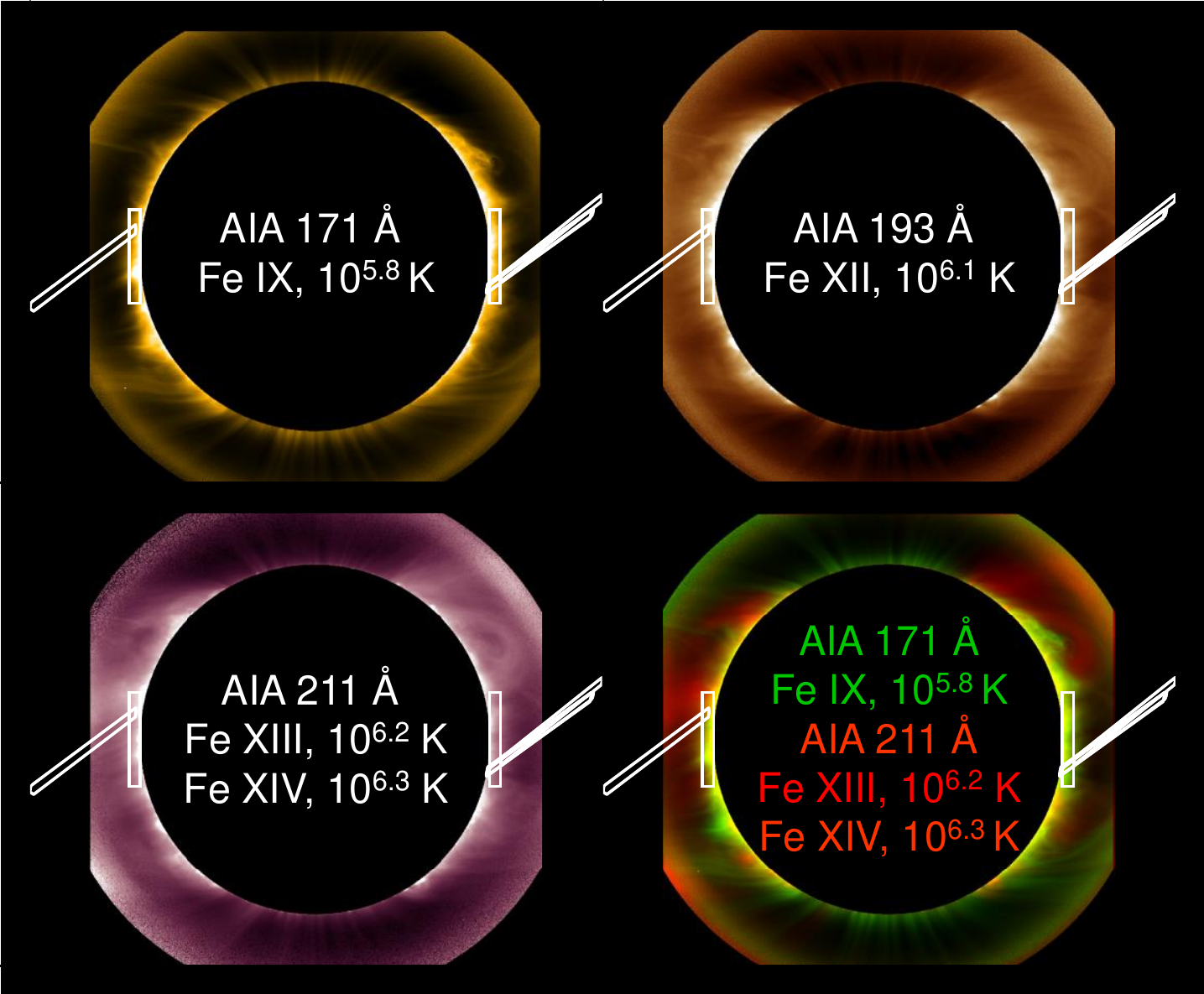}
	\caption{False-color images of the solar corona during the 2019 total
          eclipse in three SDO AIA channels.  The bottom right image is a composite obtained from AIA 171~\AA\ (green) and 211~\AA\ (red). The AIR-Spec slits and EIS fields of view are shown in white.}
	\label{fig:aia_overview}
\end{figure*}

\begin{figure*}[!htbp]
	\centering
	\includegraphics[angle=0,width=0.85\linewidth]{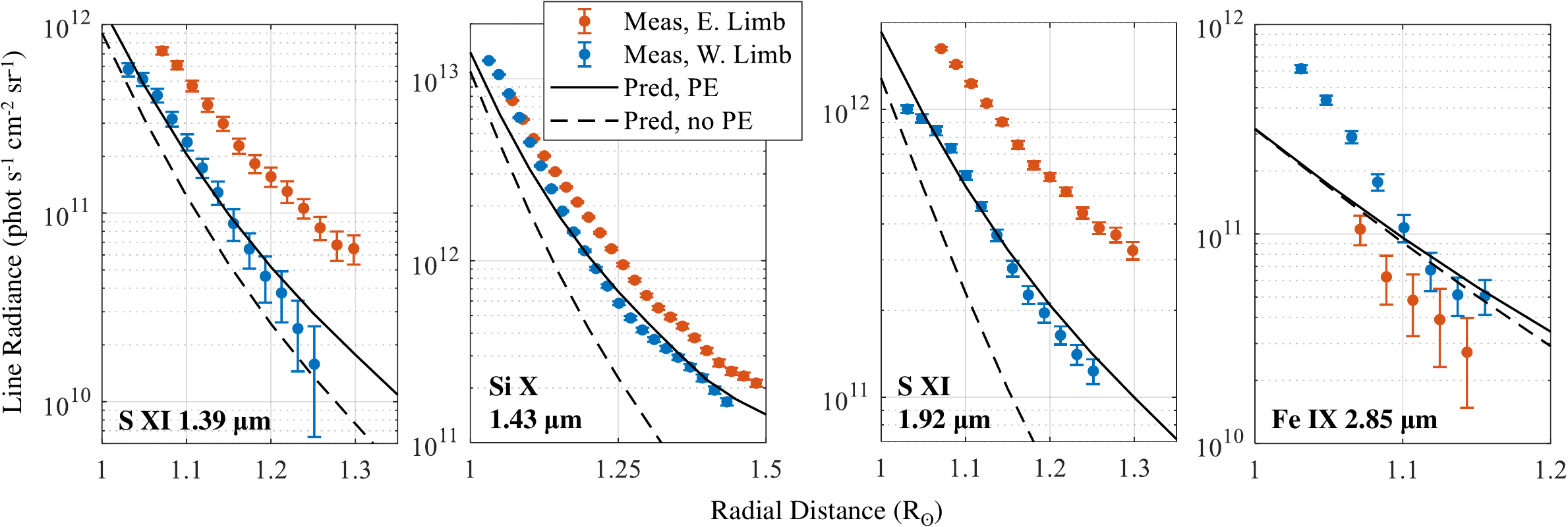}
	\caption{Predicted and measured radiance of the main four AIR-Spec lines  above the east and west limbs.  The hotter \ion{S}{11} lines are stronger above the east limb while the cooler \ion{Fe}{9} line is stronger above the west limb. Predictions with (full line) and without (dashed line) photoexcitation were made using the
      quiescent streamers density model of \cite{delzanna_etal:2018_cosie}   and 
      an isothermal temperature of 1.3 MK. This model fits the west region best,
      but does not represent well the east region. 
 }
	\label{fig:air_spec_radiances}
\end{figure*}

\begin{figure*}[!htbp]
	\centering
	\includegraphics[angle=0,width=0.7\linewidth]{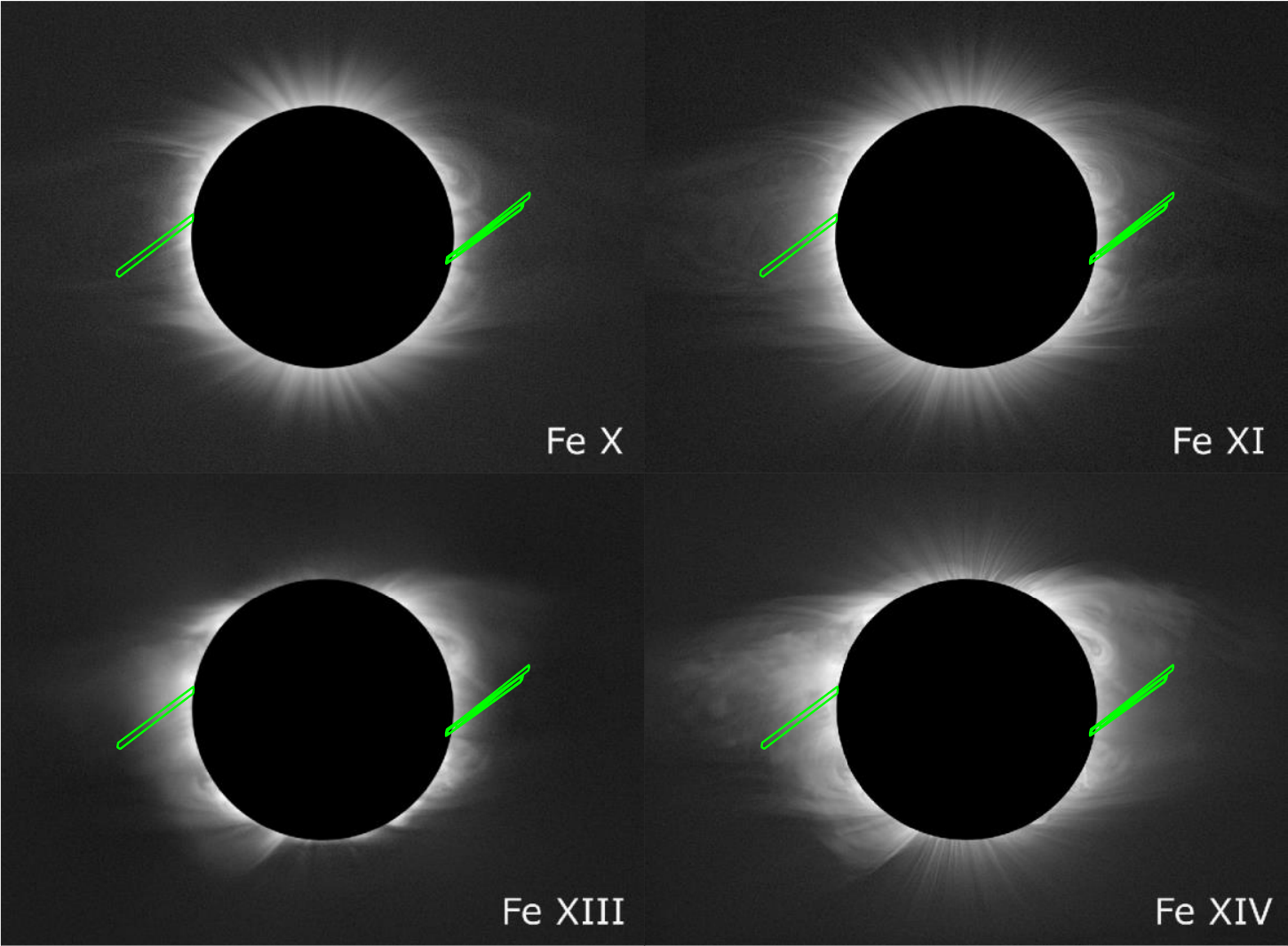}
	\caption{Narrowband images of the solar corona during the 2019 total
          eclipse in \ion{Fe}{10} 637.4 nm, \ion{Fe}{11} 789.2 nm, \ion{Fe}{13} 1074.7 nm, and \ion{Fe}{14} 530.3 nm (private communication S. Habbal to J. Samra, image credit S. Habbal and M. Druckm\"uller).   The AIR-Spec slits are overlaid in green.}
	\label{fig:vis_lines}
\end{figure*}

As we did in 2017, we coordinated multi-wavelength observations for the 2019 eclipse,
in particular with the Hinode EIS instrument and the ground-based PolarCam,
a novel micro-polarizer camera.
The effectiveness of this camera for solar observing was demonstrated during the 2017 total solar eclipse, 
where polarized brightness (pB)
measurements  
showed good agreement with the pB observations
of the dedicated solar observatory COSMO K-Coronagraph (K-Cor) in Mauna Loa, Hawaii \citep{judge_etal:2019}.
For the 2019 eclipse described here, the High Altitude Observatory (HAO) in Boulder organised an
expedition to the  Cerro Tololo Inter-American Observatory (CTIO) in Chile, taking 
polarization measurements in white light with PolarCam.

We note that other studies of the 2019 eclipse, based on observations in the visible, have been published. \cite{boe_etal:2022} presented narrowband
  images, and  \cite{hanaoka_etal:2021} used pB continuum images.
In this paper, we present an analysis
of the AIR-Spec data, to obtain measurements of electron densities, ionization temperatures,
and elemental abundances. We validate and compare these 
results to those obtained from 
Hinode EIS, PolarCam, and SDO AIA observations.

\section{Eclipse observations}
\subsection{AIR-Spec}

AIR-Spec observed for about 7 minutes during totality on July 2, 2019 (19:18:16 -- 19:26:48 UT)
in 9 slit positions. Details of the observation and data processing
are  described in \cite{samra_etal:2022}. The main lines that 
were observed are listed in Table~\ref{tab:airspec-lines},  with their typical 
formation temperatures (the ranges of temperatures where these lines
can be formed are shown in the Appendix  in Figure~\ref{fig:gt}). 
We discuss here the observations on the east and west limb, which are co-spatial with the Hinode EIS observations.

\begin{deluxetable}{cccc}[!htbp]
\tablecaption{AIR-Spec target lines. }
\label{tab:airspec-lines}
\tablehead{\colhead{Ion} & \colhead{$\lambda$ (nm)} & \colhead{T (MK)} & \colhead{Transition (Upper $\rightarrow$ Lower)} } 
\startdata
\ion{S}{11} & 1393 & 1.9 & 2s$^2$2p$^2$ \term {3}{P} {}{2} $\rightarrow$ 2s$^2$2p$^2$ \term {3}{P} {}{1} \\
\ion{Si}{10} & 1431 & 1.3 & 2s$^2$2p \term {2}{P} {o}{\sfrac{3}{2}} $\rightarrow$ 2s$^2$2p \term {2}{P} {o}{\sfrac{1}{2}} \\
\ion{S}{11} & 1921 & 1.9 & 2s$^2$2p$^2$ \term {3}{P} {}{1} $\rightarrow$ 2s$^2$2p$^2$ \term {3}{P} {}{0} \\
\ion{Fe}{9} & 2853 & 0.8 & 3s$^2$3p$^5$3d \term {3}{F} {o}{3} $\rightarrow$ 3s$^2$3p$^5$3d \term {3}{F} {o}{4} \\
\enddata
\end{deluxetable}

Figure~\ref{fig:aia_overview} shows a false-color SDO AIA image
 obtained from the  171~\AA\ (green) and 211~\AA\ (red) channels,
and the locations of the AIR-Spec  east and west slits, as well as the 
pointings of the EIS slit close to the limb.
It is well established that in the outer corona the 171~\AA\
broad-band is dominated by \ion{Fe}{9}, formed around 1 MK, while the
211~\AA\ has a significant contribution from \ion{Fe}{14}, formed around 2 MK, 
together with contributions from several other coronal lines
formed at lower temperatures such as those from \ion{Fe}{13}.
It is therefore already clear from Figure~\ref{fig:aia_overview} that the corona
in the east limb is hotter than that on the west. 
We will return to this issue below.

The actual pointing of the AIR-Spec slit varied in time during the eclipse.
The precise pointing of each exposure was obtained using the context images from the
white light slit-jaw camera.
As the solar corona was quiescent, exposures were averaged and data
spatially binned to increase the signal to noise.

We obtained measurements of the radiances 
of \ion{S}{11}  1.393 and 1.921 \mic,
as well as the strong \ion{Si}{10} 1.431 \mic\ and the weaker
\ion{Fe}{9} 2.853 \mic\ above the east and west limb, as shown in 
Figure~\ref{fig:air_spec_radiances}.
It is clear that the corona in the west and east is similar
in \ion{Si}{10}, which is formed around 1.3 MK, while it is significantly different
as seen in \ion{S}{11} which is normally formed at higher temperatures.
In principle, the differences in the two regions could also be due to
different Si/S relative abundances, although we will show below that
the difference is actually due to higher temperatures being present in the
east.

Figure~\ref{fig:air_spec_radiances} also 
shows the values predicted from the  quiescent streamers
density model of \cite{delzanna_etal:2018_cosie}, photospheric 
abundances and an isothermal temperature 
of 1.3 MK, on the basis of the measurements 
discussed below. It is clear that such a model represents quite well the  radiances in the west region.
We do not attempt to model the east region as it is 
multi-thermal.

\textbf{In Figure \ref{fig:vis_lines}, we show the AIR-Spec slits overlaid on the narrowband images described in \cite{boe_etal:2022}. The narrowband images are credited to S. Habbal and M. Druckm\"uller and were provided to J. Samra in a private communication from S. Habbal. It is clear from the \ion{Fe}{13} and \ion{Fe}{14} overlays that AIR-Spec sampled hotter material on the east limb than on the west, which agrees with the electron temperature inferred in \cite{boe_etal:2022}. Section \ref{sec:temp} provides quantitative support for this conclusion.}

\subsection{EIS}

Hinode EIS carried out two main `rasters', where the 2\arcsec\ slit 
scanned a 60\arcsec\ off-limb region with 120 s exposure times,
resulting in a total duration of about one hour.
The first raster  started at 18:36:42 UT above the east limb,
and ended at 19:37:37, just after the end of totality.
The second raster started shortly after the end of totality,
at 20:15:41 UT, and scanned a region above the west limb.
The full EIS spectral range, i.e. all of the short (SW) and long (LW)
wavelength channels were telemetered.
The maximum number of detector pixels along the slit, 512, were telemetered.

The EIS data were processed with custom-written 
software  \citep[see, e.g. ][]{delzanna_etal:2011_aia} which
largely follows the standard SolarSoft {\sc eis\_prep} program.
The hot and warm pixels (i.e. pixels with anomalous
  counts), and those affected by 
particle events  are flagged as `missing'. We used the SolarSoft
database for the hot and warm pixels.
One difference with the
standard software is that the missing
pixels are then replaced with  values  interpolated along different  directions,
and the results are visually inspected.
Despite this procedure, the spectra are very noisy. First, this is because of the
low signal. Second, this is because the long exposures resulted in a large number
of cosmic ray hits. Third, this is because the EIS detector has degraded
so much in the past few years that a large area is
affected by warm pixels, only some of which are included in the database.
Other differences  with the
standard software are that the spectra are rotated, to remove the
geometrical slant, and the fitting of the lines is carried out on the
spectra in data numbers (DN), fitting the bias with a polynomial.
To further proceed with the EIS analysis, we encountered two major problems,
concerning the radiometric calibration and the pointing, as described below.

\subsubsection{EIS  radiometric calibration }

An accurate radiometric calibration, at least in
relative terms, is necessary to measure all the key plasma parameters.
For example, the best measurements for the electron density 
for the present observations are obtained from the ratio of the \ion{Fe}{12} 186.8
and 192.4~\AA\ lines, as other diagnostic  lines
e.g.\ from \ion{Fe}{13}  were very weak.
The relative calibration at 186.8 and 192.4~\AA\ is not well known
and has been changing over time.
\cite{delzanna:13_eis_calib} presented a significant revision of the ground
calibration which only applied to data up to 2012.
Indications of time-dependent changes were also found, a problem
later confirmed by \cite{warren_etal:2014}.
Warren et al.\ analysed some observations of the quiet Sun off-limb,
and imposing  isothermality obtained a relative calibration for data until
2014.
We have extended this approach and performed a DEM analysis of a
quiet Sun off-limb observations close in time, 
on 2019-06-28  at 09:38 UT, with the 2\arcsec\ slit 
and 60 s exposures. The relative calibration was obtained by matching the 
observed radiances to the predicted ones, using CHIANTI version 10 
\citep{chianti_v10}.

The established relative calibration for the
short-wavelength (SW) channel was then used to cross-calibrate the
EIS spectra against  simultaneous
SDO AIA 193~\AA\ observations, taking into account the different
spatio-temporal resolutions,  following the methods
described in \cite{delzanna_etal:2011_aia,delzanna:2013_multithermal}.
In turn, the AIA calibration relies on the SDO EVE calibration,
although we note that this could be done only until 2014.
The AIA degradation in 2019 was extrapolated from the
cross-calibration EVE sounding rocket flown in 2018, although
that is currently being revised taking into account the latest
EVE sounding rocket flown in Sept 2021.

Details are provided in the Appendix, in Section~\ref{sec:eis_cal}.
The absolute and relative calibrations have typical uncertainties of 
20\%. The above methods have been extended to provide an EIS calibration
for the entire mission, which is the subject of another paper.

\subsubsection{EIS Pointing }

The precise pointing of the EIS instrument is generally not known.
The EIS  pointing is relatively
stable, except a 2--3\arcsec\ jittering on fast time scales, probably
caused by  small flexures inside the long (3 mt) structure
holding the optics.
Usually, the pointing is established by a cross-correlation
with EUV images. 
As the off-limb observations considered here do not present
any clear features, the only way we could estimate the
pointing was to compare EIS with the AIA 193~\AA\ data, 
as described in the Appendix, in Section~\ref{sec:eis_aia}.

\subsection{PolarCam}

The PolarCam experiment was sited at CTIO, under the path of totality in northern Chile, at an altitude of 2200~m. While not high enough to significantly mitigate telluric absorption, the location has favourable cloud cover, sitting above the stratus layer, as well as offering the infrastructure of an existing observatory. The experiment successfully observed totality from 20:38:45 to 20:40:53~UT on 2019 July 2.

The 4D Technologies PolarCam consists of a micropolarising array placed over a CCD detector. Each superpixel has four subpixels with alternating orientations of linear polarisation: 0, 45, 90, and 135 degrees. Images of the total linear polarisation can then be constructed from these four polarisation states. Its small size and lack of moving parts make such a camera an ideal candidate for a future CubeSat mission. For the July 2019 eclipse, we mounted the camera on a Stellarvue 70~mm f/6 refracting telescope behind a bandpass filter centred at 734~nm with 46~nm FWHM. Acquired images have $1920\times 1200$ pixels, with 2.91 arcsec sampling and 70~ms exposures.

\section{Plasma characterization}

\subsection{Electron Density}

\subsubsection{Density estimates from polarized brightness measurements}

The PolarCam data were calibrated  by subtracting a dark image that was taken with the shutter closed immediately following the eclipse. They were flat fielded by taking the same exposures but using an opal glass attenuator, performed 24 hours before the eclipse to mimic conditions as closely as possible.
We also investigated the contribution of sky polarization to the measurements and found it to be negligible \citep[in contrast to the measurements of][]{hanaoka_etal:2021}. This is to be expected, given that the sky is several orders of magnitude darker during eclipse, leading to a negligible contribution to the polarization from the sky.

The calibrated results of the subpixels with varying polarization state were combined to produce an estimate of pB for the PolarCam field of view. As in the 2017 eclipse \citep{judge_etal:2019}, these pB results were compared wuth those of K-Cor and found to be in good agreement. This comparison also gives a verification of the PolarCam plate scale and rotation angle, which were determined independently. The plate scale was determined by measuring the limb of the Moon and comparaing relative sizes of Sun and Moon for the ephemerides of CTIO.\footnote{\url{https://ssd.jpl.nasa.gov/horizons/}}
 Solar north was determined by measuring the locations of Bailey's Beads at second and third contact and using a Watts Chart \citep{watts63} to rotate our image. These charts of the lunar limb show the location of irregularities from a smooth sphere, allowing us to correlate the position of the bright Bailey's Beads with lunar (and hence solar) north. This method of determining the rotation angle of the PolarCam with respect to solar north was found to be consistent with a cross-correlation of the data with K-Cor.

\begin{figure}[!htbp]
	\centering
	\includegraphics[angle=0,width=1\linewidth]{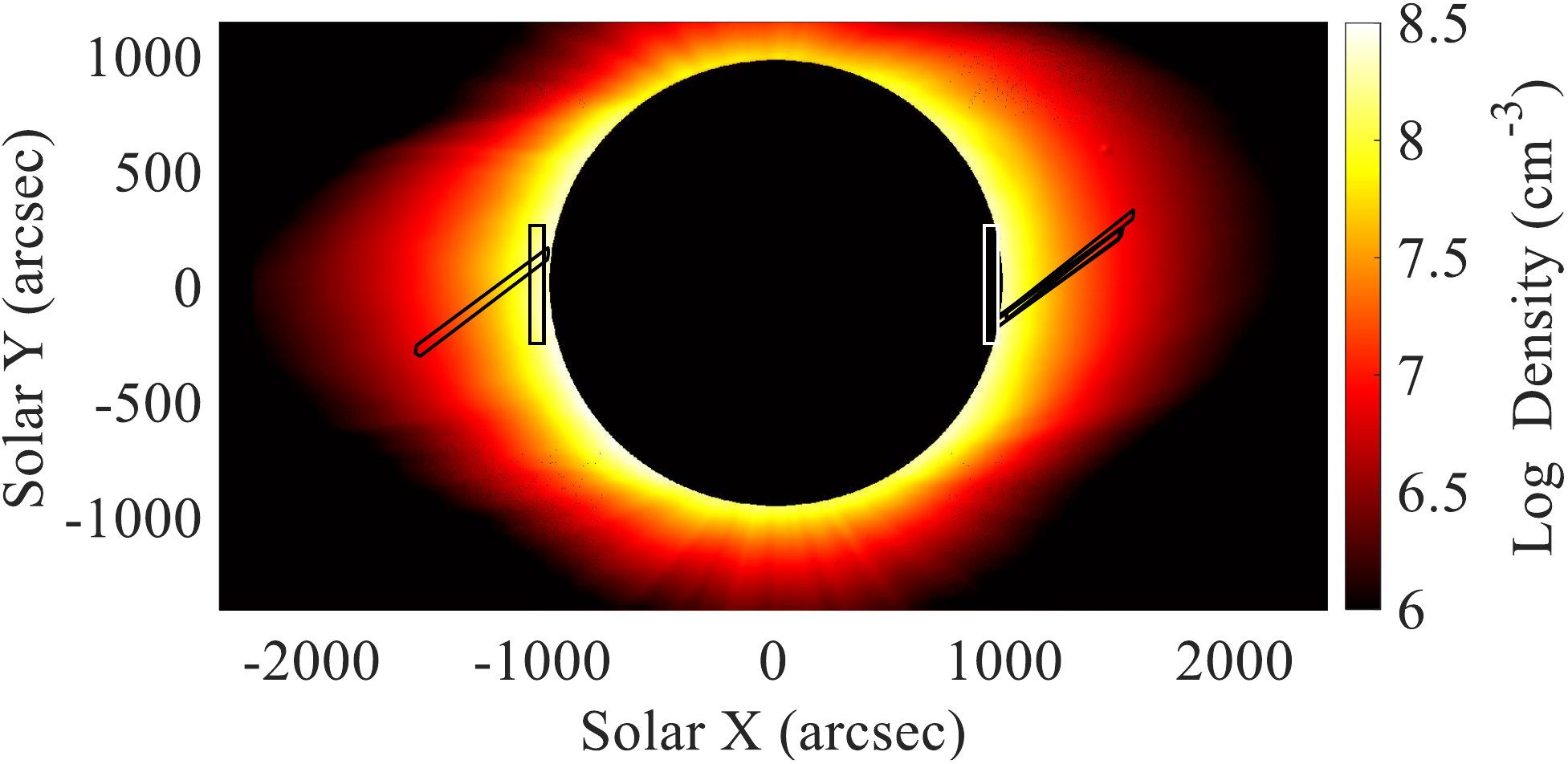}
	\caption{Electron densities obtained from the HAO PolarCam  during the 2019 total eclipse. The AIR-Spec slits and EIS FOV are overlaid.}
	\label{fig:dens_polcam}
\end{figure}

A 60~s average of the pB measurements, from 20:39:00 to 20:39:59~UT, was used to infer the electron density. 
We followed the standard assumption of 
spherical symmetry for the density 
distribution of the diffuse outer quiescent corona, and 
obtained the radial variation of the electron density
by inversion of white-light pB measurements 
following \cite{vandehulst50}.
The results are shown in Figure~\ref{fig:dens_polcam}.

\subsubsection{Density estimates from IR and EUV line ratios}

Line-of-sight (LOS) averaged electron densities can be obtained
directly from the AIR-Spec  \ion{S}{11} ratio, at least out to radial distances where
there is enough signal in the weaker line.
However, these density measurements depend significantly on the
distribution of densities along the LOS. In fact, the intensities of the
forbidden lines in the NIR are significantly pumped by PE 
from the solar disk radiation, and not just excited by electron collisional excitation (CE) and cascades.
Therefore,  the observed intensities are only
partially directly proportional to the square of the local 
electron density, as the allowed transitions are. Closer to the
Sun, CE dominates over PE, but further out, depending on the
local density, PE becomes dominant.
In other words, it is impossible to obtain accurate electron densities
from the forbidden lines unless a model for the local density distribution
is assumed. More details on this issue are presented in the Appendix,
 in Section~\ref{sec:airspec_ne}.

The obvious choice for the diffuse outer corona
is spherical symmetry, also to get consistency with the
analysis of the PolarCam observations.
We used this assumption  to model the above-mentioned
radiances of the SoHO SUMER and UVCS lines
\citep{delzanna_etal:2018_cosie}, and to obtain densities from the
\ion{Fe}{13} 1~\mic\ NIR forbidden lines observed by CoMP 
to obtain estimates of the coronal magnetic field \citep{yang_etal:2020}.

Clearly, if the density distribution is not uniform,
the densities obtained from the forbidden lines would be different to
those obtained from the allowed transitions. In general, the
forbidden lines have a significant contribution from very large
(low-density) distances along the LOS, while the intensity
of the allowed transitions mostly originates from the highest
density regions. 
A comparison of the densities obtained from these two diagnostics therefore
provides  useful information on the density distribution.
Such a comparison was carried out by  \cite{dudik_etal:2021}, using the
\ion{Fe}{13} NIR forbidden lines observed by CoMP and the
Hinode EIS  allowed transitions  from \ion{Fe}{13} and \ion{Fe}{12}.
Very good agreement in the LOS-averaged
densities from the NIR and EUV lines  was obtained in an active region
loop structure, once the background/foreground emission
of the diffuse corona was subtracted. 

It is worth noting that densities from allowed transitions are not
always independent of PE effects either.
In fact, the populations of the states in the ground configuration,
which are key to obtaining densities also from the allowed transitions,
can be affected by the solar disk radiation.
This does not occur in \ion{Fe}{12}, as most of the
transitions in the ground configuration are around 2000--3000 \AA,
where there are not many photons from the disk,
but it is not an insignificant effect in \ion{Fe}{13},
as the pumping of the  1~\mic\ NIR forbidden lines changes the
population of the ground configuration, especially the ground state.
This changes the intensities of all the lines, even those decaying to the ground state, such
as the strong 202~\AA\ line.

On a positive note, as shown 
by \cite{delzanna_deluca:2018}, the PE disk radiation can easily
be included within the CHIANTI programs assuming a 
black-body radiation at 6100 K, as it is close to 
direct 1.4--2.5~\mic\  NIR measurements from e.g. SORCE.

\begin{figure*}[!htbp]
	\centering
	\includegraphics[angle=0,width=0.99\linewidth]{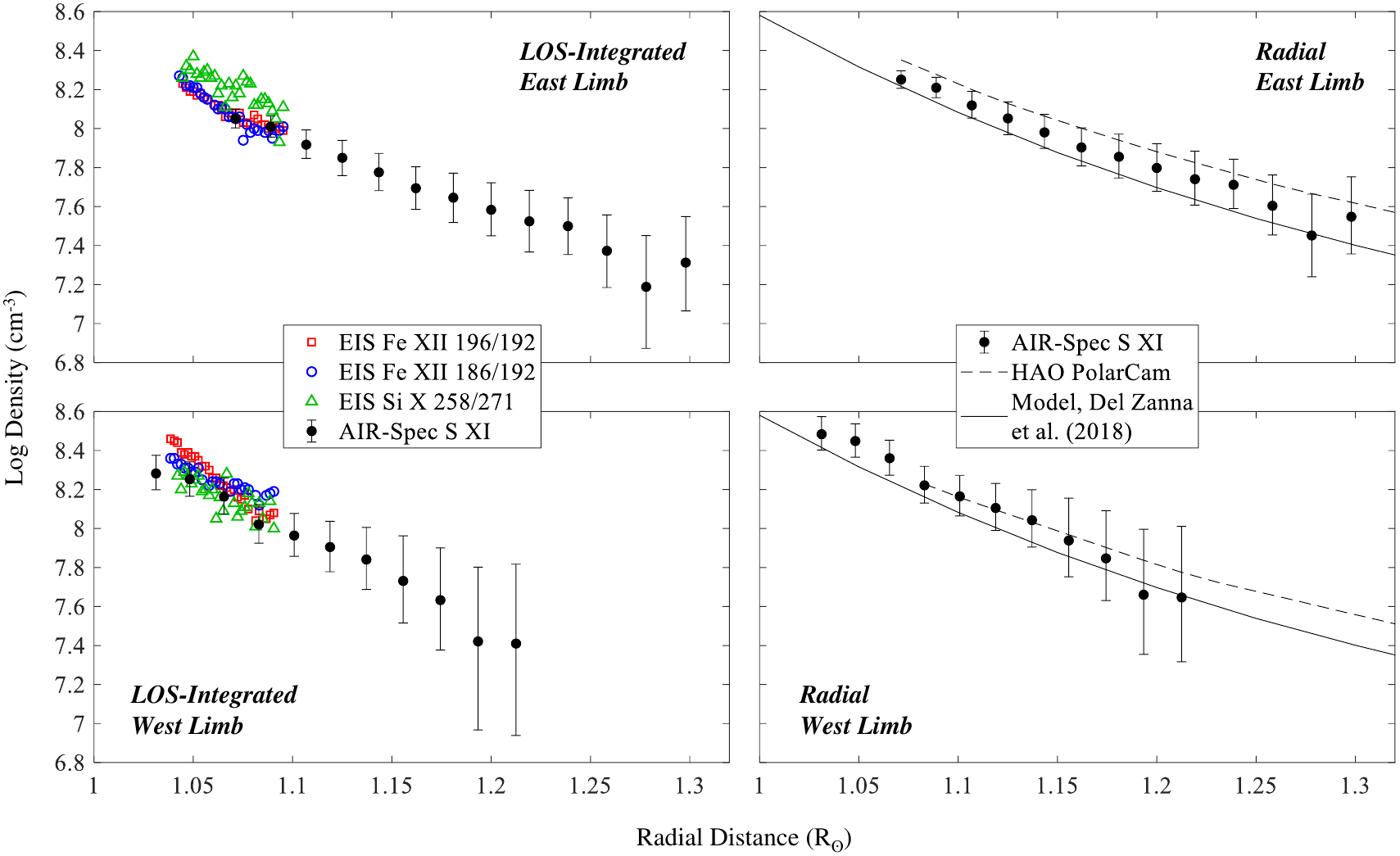}
	\caption{Electron densities on the east and west limbs during the 2019 total eclipse. The left-hand plots compare densities obtained from EIS and AIR-Spec without correcting for LOS effects. The right-hand plots compare radial density measurements obtained from AIR-Spec and PolarCam with the radial density model in \citet{delzanna_etal:2018_cosie}.}
	\label{fig:density1}
\end{figure*}

For EIS, we have 
selected the regions observed by the AIR-Spec slit,
obtained averaged radiances and densities using CHIANTI version 10
from three line ratios.
Figure~\ref{fig:density1} (left plots) shows a summary of the densities for the 
east and west. The densities from the three ratios are in agreement,
within the uncertainties (see the Appendix,  Figure~\ref{fig:eis_ne}).
The plots also show the densities from the AIR-Spec line ratio.
Again, we see very good agreement, despite the fact that the 
observations were not strictly simultaneous.
 This is to be expected as little variability was present.
The AIR-Spec densities were obtained including PE.

As mentioned, the actual radial densities would be 
slightly higher, as the line-of-sight are averaged values.
Figure~\ref{fig:density1} (right plots) shows 
the radial densities obtained from the  AIR-Spec 
measurements assuming spherical symmetry, as well as those 
from the PolarCam  observations. The results are in 
agreement, within uncertainties, and close to the 
 quiescent streamers radial density model by \citet{delzanna_etal:2018_cosie}.

\subsection{Emission measure and  relative elemental abundances from EIS}

We have taken averaged EIS spectra  in the east and west regions also
observed by Air-Spec, around 0.09~\rsun\ above the limb.
Parts of the EIS spectra are  shown in the Appendix,  in Figure~\ref{fig:eis_sp_airspec}.
The strongest lines are from \ion{Fe}{10}--\ion{Fe}{13},
 \ion{Si}{10} and \ion{S}{10}. It is clear from the spectra that the 
 east region had significant emission in the hotter \ion{Fe}{13}
 and \ion{Fe}{14}, and also \ion{Fe}{15}, which was not visible 
 above the west limb.

\begin{figure}[!htbp]
		\includegraphics[angle=-90,width=0.95\linewidth]{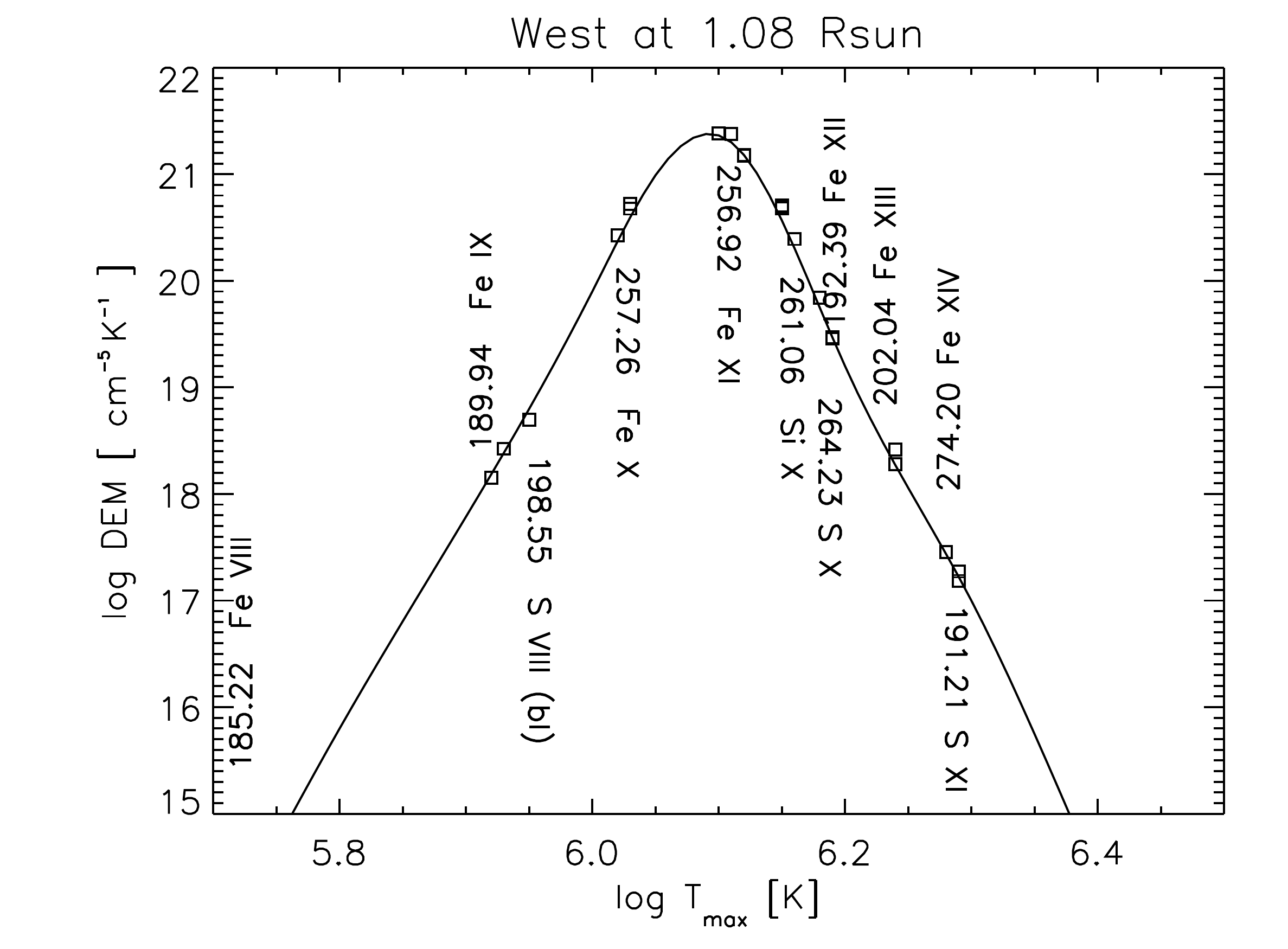}
	\includegraphics[angle=-90,width=0.95\linewidth]{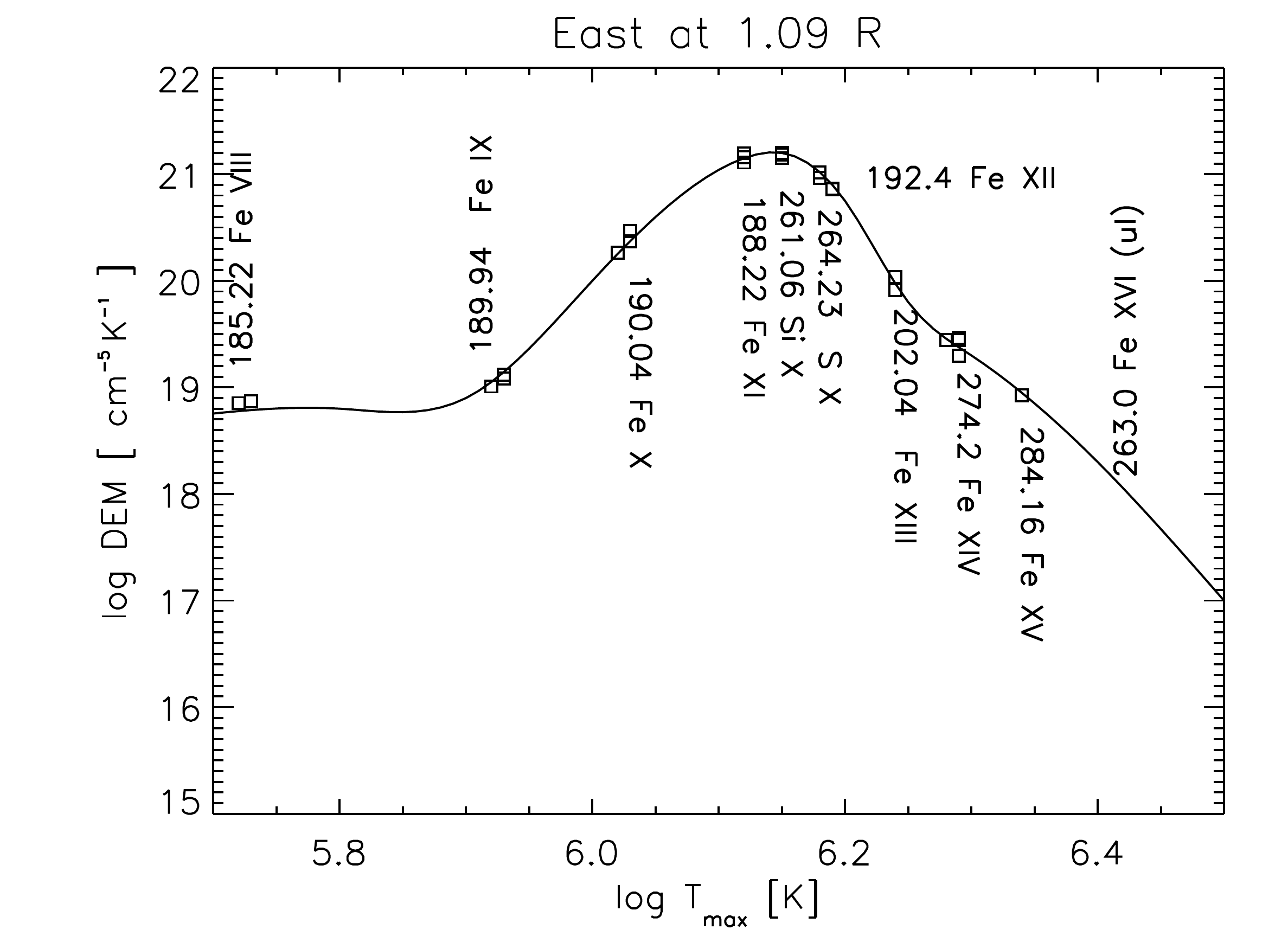}
	\caption{The DEMs for the off-limb west and east regions  obtained from Hinode EIS.
          The points are plotted at the temperature $T_{\rm max}$
          of maximum emissivity, 
 and at the theoretical vs. the observed
 intensity ratio multiplied by the DEM value. The wavelength (\AA)
      and ions of a few key lines are shown.}
	\label{fig:eis_dem}
\end{figure}

We estimated the differential emission measure (DEM)
using a modified version of the
CHIANTI\_DEM routine, where the DEM is modelled with a spline.
We used CHIANTI v.10, and to calculate the line emissivities
assumed a constant density of
1.5 10$^8$ cm$^{-3}$ (although the main lines have contribution functions that are independent of density).
We assumed the photospheric abundances of \cite{asplund_etal:2009}.
Figure~\ref{fig:eis_dem} shows the results, while more details on the lines are given in the Appendix,
in Section~\ref{sec:eis_temp}.
It is clear that the distribution is nearly isothermal above the west limb,
with a peak around 1.3 MK, while the distribution is more multi-thermal
and shifted towards higher temperatures on the east region.

The sulphur abundance is well constrained
by the \ion{S}{10} 264.2~\AA\ line. Other weaker lines from
\ion{S}{10} and \ion{S}{11} are also present and consistent with  S/Si and S/Fe
photospheric abundances, as shown in the Appendix in Section~\ref{sec:eis_temp}.
\ion{S}{10} is formed at the same temperature as \ion{Si}{10} and \ion{Fe}{12},
so the relative S/Si or S/Fe abundance is well constrained.
As sulphur, in remote-sensing observations, behaves like a high-FIP element as do Ne and Ar,
as described in the Appendix~\ref{sec:abund}, 
this is a strong indication that the quiescent streamers
had photospheric abundances, in agreement with previous observations, as reviewed in the Appendix~\ref{sec:abund}. 
AIR-Spec and AIA observations are consistent 
with these results, as discussed below. 

\subsection{Temperatures}
\label{sec:temp}

The electron temperature is a key parameter seldom measured in the 
corona. We refer the reader to  \cite{delzanna_mason:2018}
and to the brief review in Appendix~\ref{sec:isothermal}.
What we measure in this paper is the ionization temperature, i.e. 
the electron temperature from emission measure or line ratio
analyses, assuming that the quiet corona is in ionization equilibrium.
Such an assumption is reasonable for the quiescent streamers,
and has been validated with direct measurements of the 
electron temperature near the limb.

As we describe in the Appendix, there is evidence in the 
literature that the temperature distribution also in off-limb 
QS regions and in quiescent streamers, up to 3~\rsun, is 
nearly isothermal, with values around 1.3--1.4 MK. 
We have seen in the previous section that the Hinode EIS 
observations do indicate a near-isothermal distribution
above the west limb, at 1.08~\rsun. Therefore, the use of 
line (or imaging) ratios to measure `isothermal temperatures' is justified.
Clearly, if the plasma is strictly isothermal
one would expect to find the same isothermal temperatures 
from the various ratios. 

As pointed out in the Appendix~\ref{sec:isothermal},
there is also evidence in previous literature that, 
except very close to the solar limb,
the temperature is almost constant with distance, at least up 
to 3~\rsun\ (still assuming ionization equilibrium, but 
without validation).

\subsubsection{Isothermal Temperatures from AIA, EIS
and AIR-Spec line ratios, and S/Si relative abundance from AIR-Spec}

We have calculated the AIA responses as a function of temperature 
for the main coronal bands, at 171, 193 and 211~\AA, using 
CHIANTI v10 \citep{chianti_v10} atomic data and programs and the 
available AIA effective areas, obtained from SolarSoft with the
{\sc evenorm} option (in this way the AIA areas are scaled to agree with EVE),
and a constant electron density of 10$^8$ cm$^{-3}$.
We used the photospheric abundances of \cite{asplund_etal:2009},
but note that the AIA coronal bands are all dominated by iron lines
\citep[see, e.g.][]{delzanna_etal:2011_aia},
so the isothermal temperatures are independent of the choice of elemental
abundances. The AIA 193 vs 171~\AA\ and 211 vs. 171~\AA\ ratios are
strongly dependent on the ionization temperature, while the 211 vs. 193~\AA\
is not sensitive to temperatures around 1.4 MK, being flat and multi-valued.

We downloaded the AIA  data with a cadence of 2 minutes,
during the two hours of the EIS observations. There
are only minor changes in the solar corona during the
EIS rasters. However, 
for each EIS exposure, we took the AIA images closest in time
and obtained averaged AIA DN/s in the two bands for the EIS 
regions co-spatial with the  locations of the AIR-Spec  slits.

\begin{figure}[!htbp]
	\centering
	\includegraphics[angle=0,width=0.98\linewidth]{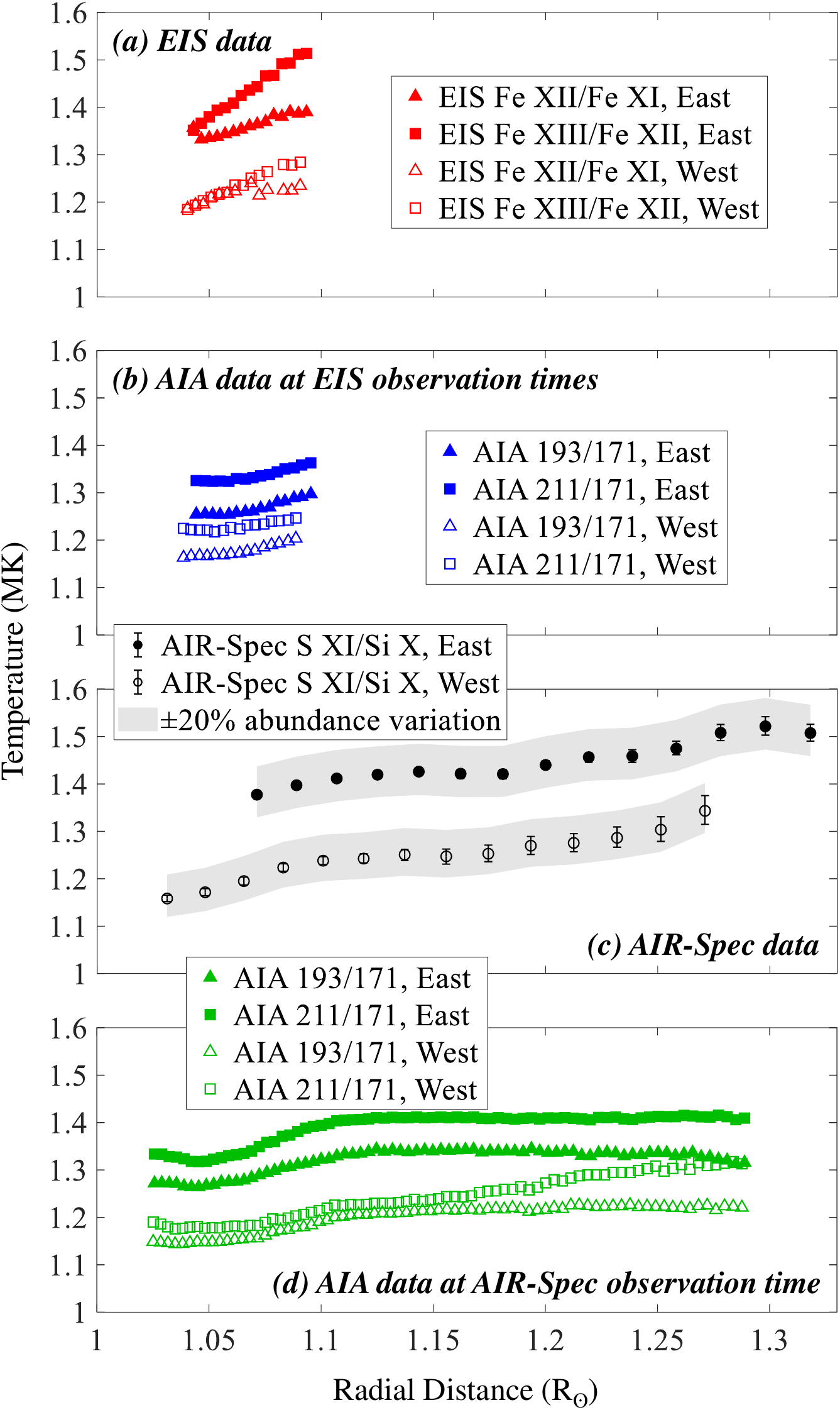}
	\caption{Isothermal temperatures in the east and west  quiescent streamers
          regions covered by the AIR-Spec slits. The top two plots show temperatures obtained from simultaneous EIS (a) and AIA (b) observations. The bottom two plots show temperatures obtained from simultaneous AIR-Spec (c) and AIA (d) observations.}
	\label{fig:temps_by_instrument}
\end{figure}

Figure~\ref{fig:temps_by_instrument}  (panel b) 
shows the isothermal temperatures obtained
from the AIA filter ratios. It is clear, as we have already anticipated, that
the eastern region is hotter than the western one. 
Figure~\ref{fig:temps_by_instrument}  (panel a) 
shows a few isothermal temperatures obtained from EIS iron lines.
There is generally good agreement
between the EIS and AIA results, confirming that the eastern region
is hotter and more multi-thermal than the western one, 
in agreement with the EIS DEM results.
However, we point out that the  
two AIA filter ratios produce an effective temperature biased by the 
thermal distribution and the AIA responses, as discussed below.

\begin{figure}[!htbp]
	\centering
	\includegraphics[angle=0,width=0.98\linewidth]{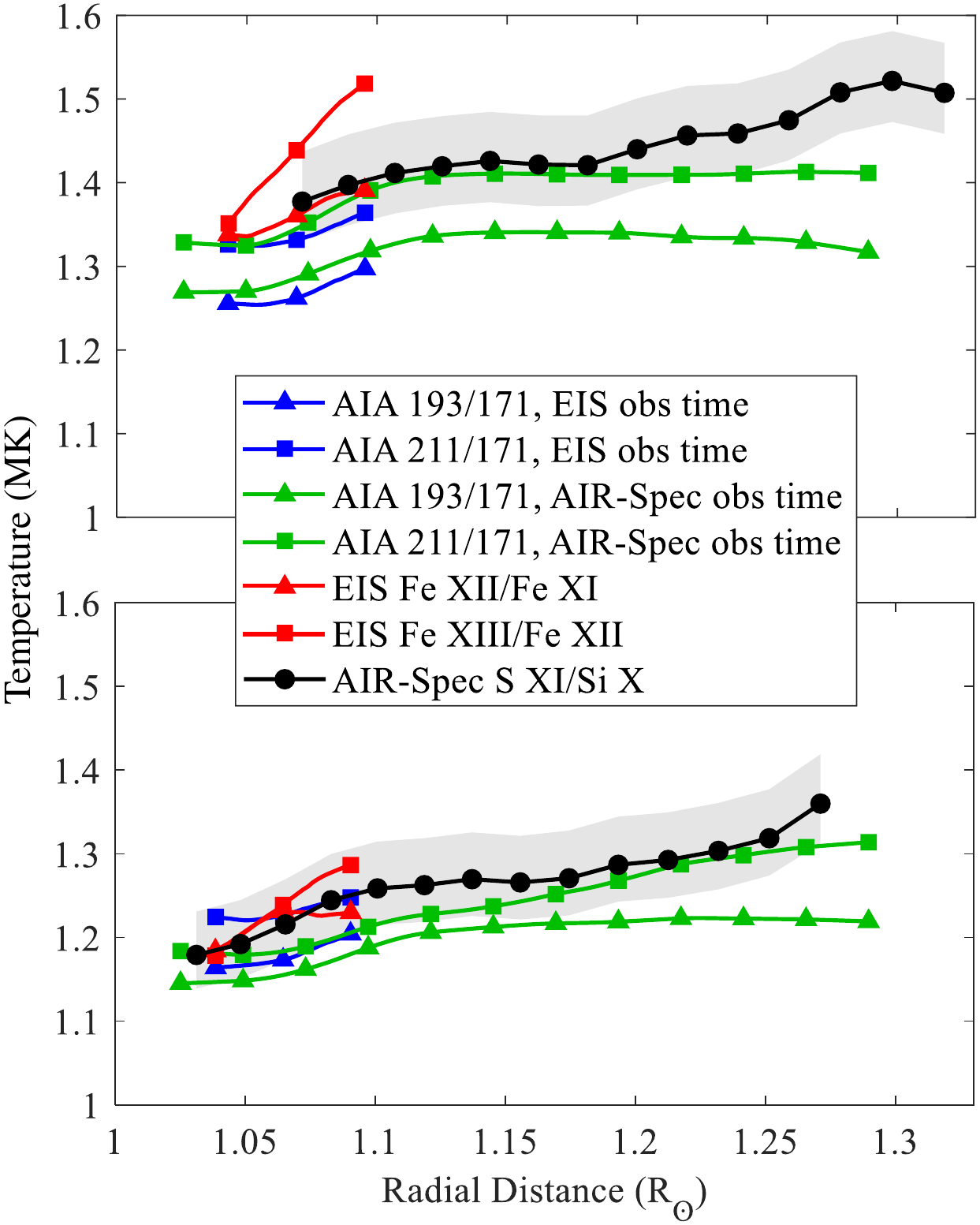}
	\caption{Overlay of temperatures from EIS, AIR-Spec and AIA observations in the east (top) and west (bottom). For clarity, the data from Figure \ref{fig:temps_by_instrument} have been smoothed before plotting.}
	\label{fig:temps_by_region}
\end{figure}

We have also calculated the AIA temperatures 
by averaging images during the 2 minutes of the total eclipse,
for a direct comparison with the AIR-Spec data. 
The results are shown in Figure~\ref{fig:temps_by_instrument}, 
bottom panel, and in Figure~\ref{fig:temps_by_region}.

Unlike the previous 2017 observation, where a few lines were observed and
enabled us to produce an emission measure loci, in the present case we only
have two AIR-Spec lines to estimate the ionization temperature:
\ion{Si}{10} 1.431 \mic\ and \ion{S}{11} 1.921 \mic\ out to 1.3 \rsun, as the
\ion{Fe}{9} 2.853 \mic\ is too weak at large distances.
The \ion{Si}{10} vs. \ion{S}{11} ratio is in principle a good temperature diagnostic  for quiescent streamers,
but there are two issues. The first regards the relative elemental abundances
of Si and S. 
 As the relative S/Si abundances as measured by EIS
at 1.08~\rsun\ are photospheric, we expect them to be the same at larger distances, so we used the same S/Si abundances
to calculate the isothermal temperature from the NIR ratio.

The second issue relates to the above-mentioned
complication that the LOS radiances
not only depend on density and temperature, but also on the relative
importance of the CE and PE processes.

We have adopted the model of spherical symmetry, photospheric abundances and the
densities we obtained from the \ion{S}{11} ratio, to predict the
\ion{Si}{10} vs. \ion{S}{11} LOS radiances as a function of temperature
(see details in the Appendix, Section~\ref{sec:airspec_ne}).
By comparison with the observed ratio of the radiances, we then
obtained the isothermal temperatures from the NIR ratio.
The values are shown as diamonds in Figure~\ref{fig:temps_by_instrument}, panel c and as circles in Figure~\ref{fig:temps_by_region}.

The gray shaded area in  Figures~\ref{fig:temps_by_instrument},\ref{fig:temps_by_region}
indicates the lower and higher temperatures
obtained from the NIR ratio by changing the relative S/Si abundance by $\pm$20\%.
As there is close agreement between the
AIR-Spec temperatures and those obtained from the AIA 211/171~\AA\
filter ratio, this shows that the S/Si relative abundance are 
constant with distance, as expected. 
In fact, the AIA temperatures are 
independent of the choice of abundances. 
If the S/Si relative abundance differed from the photospheric 
value by more than 20\%, the NIR temperatures would differ from the 
AIA (and EIS) ones.  
The predicted radiances of the NIR lines obtained form the sperical symmetry model for the west region are also shown in 
Figure~\ref{fig:air_spec_radiances}. Having established independently from the \ion{S}{11} the electron density,
it is clear that a constant temperature model reproduces quite well
the radial fall off of the NIR radiances in \ion{Si}{10} and 
\ion{S}{11} quite well. If we used relative Si/S abundances 
that were not photospheric, we would find disagreement with the 
observed radiances. 

\subsection{AIA forward-model and photospheric abundances}

\begin{figure}[!htbp]
	\centering
	\includegraphics[angle=0,width=1\linewidth]{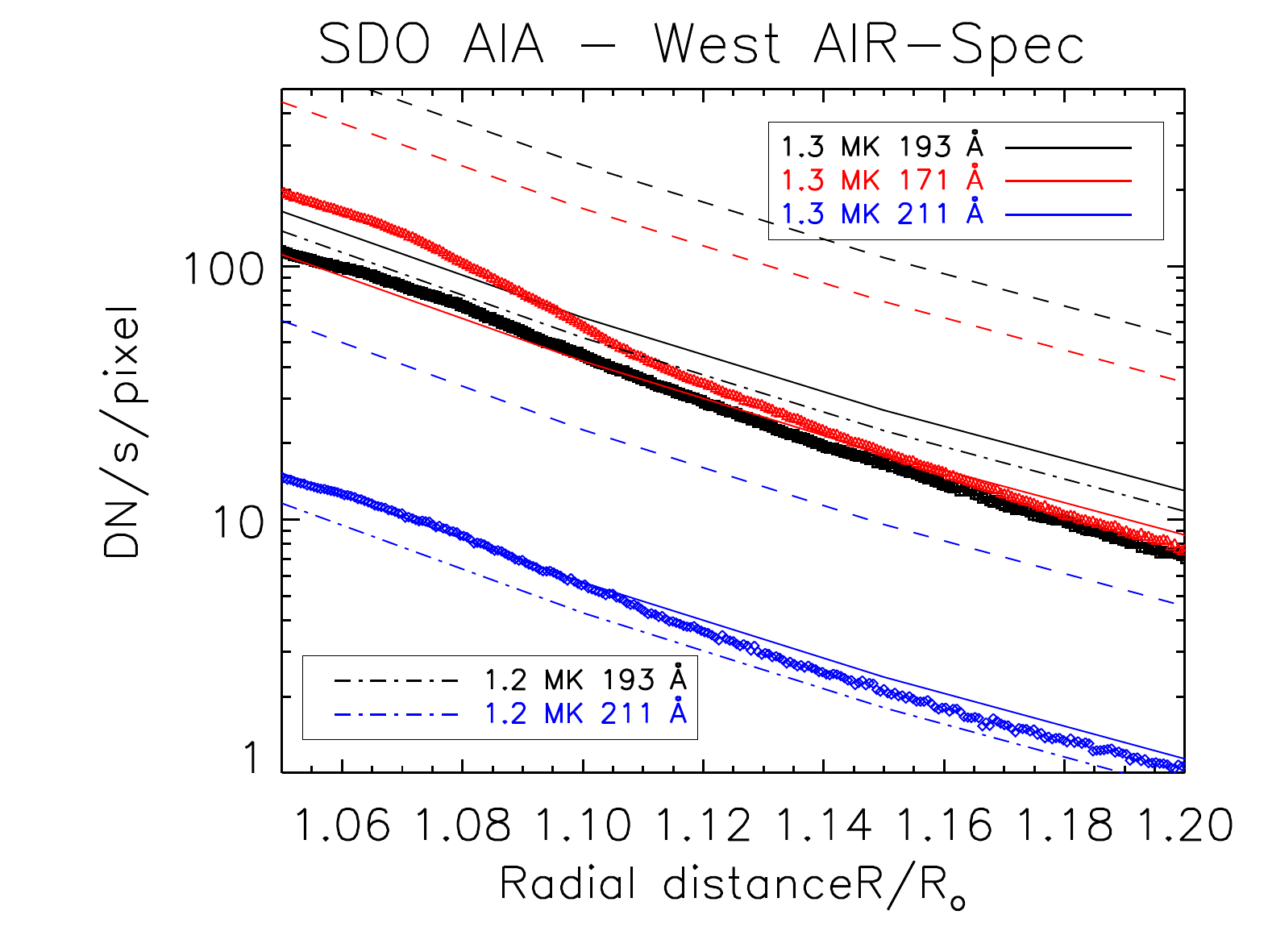}
	\caption{Measured and predicted AIA count rates in the  west region. Symbols indicate the measured count rates. Solid lines show the predicted count rates using a spherical symmetry model, photospheric abundances, and assuming an isothermal and constant temperature of 1.3 MK. Dashed lines indicate the predicted signal increased by a factor of four. Dot-dashed lines show the values with the 1.2 MK isothermal 
	model.}
	\label{fig:aia_dns}
\end{figure}

Having estimated the radial  electron density and temperature,  it is then 
straightforward  to calculate the signal in the AIA bands using the spherical model and compare it to 
observation. 
The AIA signal depends not only on the density and temperature,
but also on the absolute AIA radiometric calibration, and 
on the absolute iron abundance, as the AIA bands are dominated by Fe lines.
In quiescent streamers, 
the 171~\AA\ band is dominated by Fe IX, while the 
193 band has several Fe XII and Fe XI lines. The 211~\AA\ 
is mostly dominated by Fe XIII lines.
The absolute AIA calibration has some uncertainty, but the degradation
in these coronal bands has been relatively small (10-30\%) over 
the years, so we do not expect an uncertainty larger than say 30\%,
which is the typical uncertainty in the calibration 
of the sounding rockets used for the in-flight calibration of AIA.
Therefore, the comparison provides an estimate of the  absolute iron abundance, within 30\%.

Considering that a spherically-symmetric  isothermal model with T=1.3 MK
and the radial density profile of Del Zanna et al. (2018) 
represents the AIR-Spec  data  in the west region quite well, we have used it to 
calculate the AIA count rates up to 1.3~\rsun.
For simplicity, we have calculated only the \ion{Fe}{8} -- \ion{Fe}{14}
emissivities and folded them with the expected AIA effective areas.
In fact, these are the main ions contributing
to the bands.
The results with photospheric abundance are shown, together with the 
observed ones, in Figure~\ref{fig:aia_dns}. It is clear that 
there is an overall agreement, 
showing that such simple model can also 
represent the AIA observations, and that 
the absolute iron abundance on the west streamer must be close to photospheric. 
If the iron abundance were coronal, the observed count rates would be a factor of 4 higher. 

The small discrepancy between the isothermal temperatures
obtained from the AIA filter ratios and the assumed constant
temperature of 1.3 MK is related to the fact that the ratios we have used, 193 or 211 vs. the 171~\AA, provide 
an effective temperature weighted by the DEM distribution. 
If we take the predicted count rates and estimate the isothermal
temperature, we get 1.2 MK from both 211/171 and 193/171~\AA\ ratios,
a value close to the peak of the DEM (1.26 MK).

Clearly, the isothermal assumption is an approximation to the 
real distribution, which we estimated only at 1.08~\rsun\ with 
Hinode EIS. Table~\ref{tab:aia} shows that the AIA count rates 
as predicted with a 1.2 and 1.3 MK isothermal models and with the EIS DEM  are close, within 50\%. 
Varying the isothermal temperature by 0.1 MK does not have a large
effect on the predicted count rates, as shown in Figure~\ref{fig:aia_dns}.

\begin{deluxetable}{lccc}[!htbp]
\tablecaption{AIA observed and predicted DN/s per pixels 
in the west region at 1.08~\rsun}
\label{tab:aia}
\tablehead{\colhead{} & \colhead{171~\AA} & \colhead{193~\AA} & \colhead{211~\AA} } 
\startdata
Observed               & 103  & 69 & 8.6 \\
Predicted (1.2 MK)     & 93   & 77 & 6.4 \\
Predicted (1.3 MK)     & 62.  & 92 & 8.4 \\
Predicted (DEM)        & 55.  & 57 & 6.2 \\
\enddata
\end{deluxetable}

\section{Conclusions}

With an improved AIR-Spec instrument, we have shown here the diagnostic potential
of NIR lines to measure  reliably and independently electron temperatures, densities and  elemental
abundances in the corona, out to  large distances. 
For the 2019 eclipse, we could only measure lines up to 
1.3~\rsun, because of the limited size of the instrument and the slit,
in addition to the short exposures, which were necessary to observe large sections 
of the corona during the eclipse. 
However, NIR lines, being strongly photo-excited by the disk radiation as
the visible lines, have relatively strong radiances out to large distances,
hence are excellent probes to study the middle corona. 

The AIR-Spec results are  in agreement with  those obtained in the 
low corona by the EIS instrument. Consistency with the 
densities obtained from the PolarCam pB measurements and with  
the intensities in the SDO/AIA coronal bands is also reassuring. 

In agreement with the analysis of the first 2017 AIR-Spec flight
\citep{madsen_etal:2019},  we found that the quiescent streamers 
have Si/S photospheric abundances, from both the  AIR-Spec 
and the Hinode EIS lines. The spherical symmetry model
which reproduces quite well the AIR-Spec data indicates that 
the absolute iron abundance is close to its photospheric value,
by comparison with the AIA count rates.     
These results are also in agreement with previous 
literature (reviewed in the Appendix~\ref{sec:abund}) 
pointing out that the quiescent solar corona
during solar minimum has near photospheric abundances, at least 
up to 1.3 MK. 

Also, in agreement with previous observations and the 
2017 AIR-Spec flight, we found that the quiescent streamer above the west limb is nearly isothermal. On the other hand, the 
eastern  one had some hotter emission. The results from EIS, AIA 
and AIR-Spec are all consistent, and also in agreement with the
eclipse observations in the visible.
A simple spherical symmetry model with constant temperature
with radial distance reproduces quite well the 
AIR-Spec and AIA observations, up to 1.3~\rsun. 

We have shown the importance of 
measuring at least a few key lines formed at different temperatures 
to establish the thermal distribution of the plasma, as we have done
at lower heights with Hinode EIS. In fact, single 
isothermal ratios have some limitations, as they provide effective temperatures
which are averaged values weighted by the DEM distribution
and their temperature sensitivity.

It is also clear 
that the analysis of NIR lines is non-trivial as careful atmospheric modeling is required,
as well as a model of the 
electron density and temperature distribution, as 
the radiances cannot be directly used to measure these parameters.
This is because of the variable 
contribution of the resonantly scattered component to the line emission,
compared to the collisionally excited one. 
Once the electron density is established, it is relatively easy to 
find a temperature distribution that fits the observations. 
Therefore, we recommend that future NIR observations include at least a 
density-sensitive line ratio such as the  \ion{S}{11}, and at least 
three lines formed in the temperature range between 1 and 2 MK
for the quiet corona.

As briefly described in \cite{delzanna_deluca:2018}, the NIR 
spectral region is virtually unexplored but has a great plasma diagnostic potential. 
The follow-up to AIR-Spec is the  Airborne Coronal Emission Surveyor (ACES), a new imaging Fourier transform spectrometer that will survey the entire 1–4 \mic\ spectral range at high (0.5 cm$^{-1}$) spectral resolution during the April 8, 2024 total solar eclipse.
ACES will fly along the path of totality on the GV, with a 20 cm  solar feed stabilized to 6\arcsec\ RMS over one second. The feed will be provided by the Airborne Stabilized Platform for InfraRed Experiments (ASPIRE), which was developed by the AIR-Spec/ACES team for the GV and commissioned in 2021. ACES will measure many coronal lines, half of which have never been observed before, and will therefore be a pathfinder for future targeted DKIST and  hopefully space-based observations.

\acknowledgments 
    { The 2019 AIR-Spec upgrade and re-flight was funded by NSF award \# AGS-1822314: \textit{Airborne InfraRed Spectrograph (AIR-Spec) 2019 Eclipse Flight}.
      GDZ and HEM acknowledge support from STFC (UK) via the consolidated grants to the atomic astrophysics group (AAG) at DAMTP, University of Cambridge (ST/P000665/1. and ST/T000481/1).
      GDZ also acknowledges support from NSF award \# AGS-2117582: \textit{MRI: Development of an Airborne Coronal Emission Surveyor (ACES)} for his visit to CfA in 2022. 
 Hinode is a Japanese mission developed and launched by ISAS/JAXA, with NAOJ as a domestic partner and NASA and STFC (UK) as international partners. It is operated by these agencies in cooperation with the ESA and NSC (Norway).
 SDO data were obtained courtesy of NASA/SDO and the AIA and HMI science teams.
 The PolarCam experiment was supported by the National Center for Atmospheric Research, which is a major facility sponsored by the National Science Foundation under Cooperative Agreement No.\ 1852977.
  We thank the reviewer for the detailed comments which helped us to 
  improve the manuscript.
    }

\bibliographystyle{aasjournal}
 \bibliography{density_paper.bib}



 \appendix

\section{Elemental abundances in the quiet Sun}
\label{sec:abund}

Solar coronal abundances measured with remote-sensing
instruments vary with the structure observed and the temperature
of formation of the spectral lines observed.
For a review of measurements of different solar features we refer to
 \cite{delzanna_mason:2018}.
Historically, most observations have constrained the relative
abundances between high- (e.g. Ne, O, Ar) and low-FIP (e.g. Si, Fe, Mg) elements, and take the ratio with 
 the photospheric ones to obtain the FIP bias.
We adopt here the photospheric abundances of  \cite{asplund_etal:2009},
but note that the values for several elements (especially
C, N, O) are still debated in the literature, although
variations from Asplund et al. are of the order of 30\%.
We also note that the Ne, Ar abundances recommended
by \cite{asplund_etal:2009} are not based on
actual photospheric  measurements.

We focus here on `quiet Sun' (QS) areas during solar minimum, and note that 
the presence of active regions has a significant effect on the diffuse QS emission, as shown e.g. in
\cite{delzanna_andretta:2011,andretta_delzanna:2014}.

Early QS observations  from Skylab
in transition-region lines indicated near photospheric
relative abundances, when  considering Mg/Ne ratios
\citep[see, e.g.][and references therein]{feldman:92,feldman_widing:1993}.
Similar results were obtained with SoHO CDS, also
considering Mg/Ne ratios \cite[see, e.g.][]{delzanna_jgr99a, young:2005}.

SoHO SUMER was an excellent instrument to measure the relative abundances
of coronal lines in off-limb QS regions.
\cite{doschek_etal:1998_si_ne} used SOHO SUMER observations
to find photospheric  Si/Ne abundances in the upper transition region.
Other results involving more elements 
were unclear, as they depended on which spectral lines
were considered, as discussed by \cite{feldman_etal:98b}.
The problem was partly related to the inclusion or not 
of lines from the so-called `anomalous' ions
of the  Li-like and Na-like isoelectronic sequences
\citep[see e.g.][and references therein]{delzanna_etal:02_aumic}.

We note that the atomic data for the forbidden lines of Fe ions
in the SUMER spectra were significantly revised by one of us (GDZ),
as summarised in \cite{delzanna_mason:2018}.
The new atomic data were included in  CHIANTI version 8 \citep{delzanna_chianti_v8}.
A reanalysis of the SUMER observations with these updated atomic data
have indicated relative photospheric abundances, considering a range of low-FIP
elements such as Fe, high-FIP ones such as  Ne, O, and Ar, and the mid-FIP
S  \citep{delzanna_deluca:2018}.

The SUMER off-limb spectra were  also used by \cite{feldman_etal:98b}
to find an O/H ratio to be close to the  photospheric value. 

Hinode EIS has been used extensively in the literature to measure
the FIP bias. The instrument observes strong coronal lines from Fe and Si
(low-FIP), but much weaker lines from S and Ar.
In most observations, only a weak Ar XIV can reliably provide the
Fe,Si vs. Ar relative abundance in 3 MK plasma, found in the cores
of active regions. The results are that the FIP bias is always about a
factor of 3, as in the example shown in  \cite{delzanna:2013_multithermal}.
This agrees with an analysis of many quiescent active region
cores observed in the X-rays, where the Fe vs. O, Ne FIP bias is also about 3
\citep{delzanna_mason:2014}, i.e. lower than  4, the earlier 
widely adopted value recommended by \cite{feldman_etal:92a}.

The Hinode EIS observations of S XIII, also formed around 3 MK,
confirm an Fe, Si vs. S FIP bias of about 3 \citep{delzanna:2013_multithermal}
(while the lower charge states of S often indicate a lower FIP bias
of about 2). Therefore, S in remote-sensing observations shows the same FIP bias as the high-FIP elements O, Ne, so effectively
behaves as a high-FIP element, despite having a FIP of only 10 eV.
The behaviour of S as a high-FIP element is also observed in
solar energetic particles (SEP)
events measured in-situ, and predicted by Laming's theory
\citep[see, e.g.][]{reames:2018,laming_etal:2019}. 
It is also interesting to note that the FIP bias of SEP is 
about 3, in agreement with the recent results for active region cores.

On the basis of the above arguments, there is an extended
literature on FIP bias results obtained from Hinode EIS
measurements of the strong S X vs. Si X lines around
264~\AA, as they are formed at similar temperatures (which 
depend on the feature observed),  and are close in wavelength
(so the uncertainty in the radiometric calibration is small).
When observing the QS, the S X always indicates a relative
Si, Fe vs. S photospheric abundance, while
active regions show a FIP bias of around 2
\citep[see,e.g.][]{delzanna:2012_atlas,doschek_warren:2019}.

The AIR-Spec NIR spectra of the 2017 eclipse, when the Sun was 
very quiet, have also indicated a photospheric S/Si abundance
\citep{madsen_etal:2019}.

It could well be that QS plasma at temperatures higher than 1 MK
has a FIP bias, but past/current instrumentation is not sensitive enough to measure it. Strong lines from low-FIP elements formed 
around 2 MK are usually very weak, and lines formed around 3 MK
have no signal in past and current instrumentation. 
Spectral lines from high-FIP formed around 2 MK or more have no signal in the quiet Sun.

Regarding absolute abundances, in earlier literature it was not
clear if it was the low-FIP elements that were more abundant in 
active regions, or the high-FIP ones that were less abundant. 
Line to continuum measurements during large flares, when 
the FIP bias is generally 1, indicated photospheric abundances
for all the elements. Regarding quiescent active cores, 
on the basis of a filling factor argument, \cite{delzanna:2013_multithermal} suggested that it should be 
the  low-FIP elements that have increased abundances. 
Line-to continuum measurements in the X-rays of the full Sun,
when a single active region was dominating the signal, indicated a Si FIP bias of 2.4 \citep{delzanna_etal:2022_xsm}, i.e.
only about 20\% lower than the  value suggested by Del Zanna.

Irradiance measurements (i.e. Sun-as-a-star) have been carried out
since the 1960's in the X-rays and EUV. Most measurements were carried
out outside of solar minimum, hence the presence of active regions
affected the results in terms of elemental abundances. One example is the 1969 sounding rocket flight
described by  \cite{malinovsky_heroux:73}. It obtained the best-calibrated
EUV spectrum of the Sun, but active region lines (formed above 2 MK) and even flare lines were present.
\cite{laming_etal:1995} used that spectrum to find 
 nearly photospheric abundances  when considering 
 lines formed in the transition region and corona, up to 1 MK.
 The abundances were relative, using mainly sulphur lines, in combination with lines from  low-FIP elements.
An FIP bias of 3--4 was instead obtained from  higher-temperature lines.
A re-analysis of the same spectrum  by \cite{delzanna:2019_eve} using CHIANTI version 8
found similar results, showing no FIP bias when considering
S X, and an increase to a FIP bias of about 2 when considering S XI and S XII,
formed around 2 MK. As explained by 
\cite{delzanna:2019_eve}, this difference is simply due to the
fact that the QS  dominates the S X emission (hence the FIP bias is 1),
while active regions dominate the S XI, S XII, hence the increase
in the FIP bias. Finally, we note that X-ray bright points also contribute to the irradiances
at 2 MK with an FIP bias of about 2, as found from line to continuum measurements in the X-rays \citep{xsm_XBP_abundance_2021}.

On the other hand, 
when analysing irradiances during solar minimum, 
the relative abundance between sulphur lines and those
of low-FIP elements is photospheric, as shown by 
\cite{delzanna:2019_eve} using the well-calibrated 
spectra measured during the solar 
minimum on 2008 April 14 by the prototype of the 
Solar Dynamics Observatory Extreme ultraviolet Variability Experiment (EVE).  

\section{Near-isothermal and constant electron temperature in  quiescent streamers}
\label{sec:isothermal}

As reviewed in  \cite{delzanna_mason:2018}, measuring the electron temperature
directly from e.g. line ratios is possible but was rarely applied to
observations, in particular those of the quiet Sun corona above the limb,
as its temperature is about 1 MK and very few diagnostics are available.

The best one is a line ratio from \ion{Mg}{9}.
\cite{delzanna_etal:2008} produced significantly 
improved atomic data for this ion, and used off-limb SoHO SUMER
measurements reported by  \cite{feldman_etal:1999}  to
find a  temperature of 1.35 MK. 

Another possible diagnostic, applied to a quiescent streamer
at a large distance (2.7\rsun), involved measuring the extremely weak electron-scattered 
component of the neutral hydrogen Ly~$\alpha$ as recorded by
SoHO UVCS  \citep{fineschi_etal:1998}.
A temperature of 1.1$\pm 0.3$ MK was found.

Lacking direct diagnostics, the usual assumption to estimate
the electron temperature is to measure the ionization temperature,
i.e. the temperature obtained with line ratios or emission measure
analyses, assuming  ionization equilibrium.
Early measurements of line ratios
from the 1970's  of the inner corona of the quiet Sun have indicated 
a temperature distribution around 1.4 MK \citep[cf.][]{jordan:1971}.

Detailed emission measure analyses obtained with a range
of coronal lines observed by SoHO SUMER clearly indicated an
isothermal distribution along the line of sight for QS off-limb regions \citep{feldman_etal:1999}. 
A reanalysis of the same SUMER observations with improved atomic data,
shown in  \cite{delzanna_mason:2018}, confirms this result, with a
temperature of 1.35 MK, in exact agreement with the  \ion{Mg}{9}
result. This provides a strong indication that for the QS the
ionization temperature is a good measure of the electron temperature,
and that the CHIANTI atomic data, in particular the
 ionization/recombination rates for the ions considered,  are accurate.
  
 There is also  evidence in the literature that
 the temperature is relatively constant with radial
 distance, if one observes  quiescent streamers,
 when no active regions are present.
For example,  Skylab ATM observations of EUV coronal lines out to 1.25\rsun\  indicated a
 nearly constant ionization temperature
 \citep{mariska_withbroe:1978}.
 SoHO UVCS observations of EUV coronal  lines in
 quiescent streamers, between 1.4 and 3~\rsun, also indicated
 a constant ionization temperature of about 1.4 MK \citep{delzanna_etal:2018_cosie}. 
 Clearly, close to the solar limb the temperature decreases towards
 lower values.
 Semi-empirical models of the electron temperature in quiescent
 streamers also indicated nearly constant values \citep[see, e.g.][]{withbroe:1991}.
 Finally, narrow-band images in iron visible lines during eclipses 
 also indicate a nearly constant ionization temperature \citep[see, e.g.][]{boe_etal:2022}.

 \section{Measuring density and temperature using IR line ratios}
\label{sec:airspec_ne}

\begin{figure}[!htbp]
	\centering
	\includegraphics[angle=0,width=0.6\linewidth]{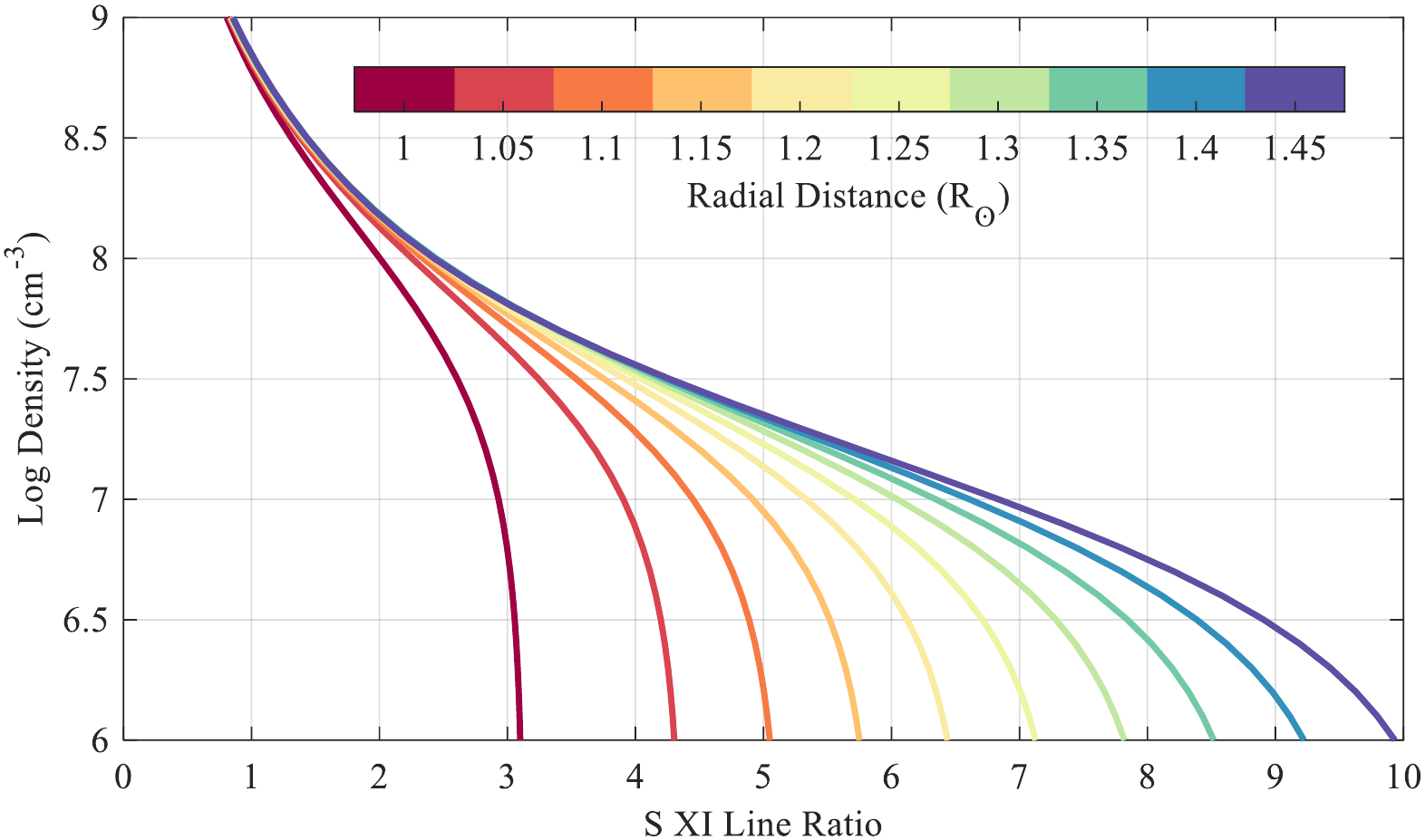}
	\caption{Density as a function of \ion{S}{11} 1.92/1.39 \mic\ line ratio and distance from Sun center.}
	\label{fig:dens_mapping}
\end{figure}

 The mapping from AIR-Spec line ratios to electron density was provided by the `DENSITY\_RATIOS' routine in CHIANTI v.10 \citep{chianti_v10}.
 As the IR lines are significantly photo-excited, the distance from the underlying photospheric radiation field is a critical parameter in the mapping: as the radiation becomes diluted with increasing distance, the pumping rate also
 decreases.   Figure \ref{fig:dens_mapping} shows the dependence of density on the combination of the \ion{S}{11} 1.92/1.39 \mic\ line ratio and the distance from Sun center. The AIR-Spec density estimates were computed by interpolating the density surface in Figure \ref{fig:dens_mapping} to each line ratio/radial distance pair in the AIR-Spec data. This procedure produces the density estimates in the left half of Figure \ref{fig:density1}, which are not corrected for LOS contributions to the AIR-Spec radiances.

\begin{figure}[!htbp]
	\centering
	\includegraphics[angle=0,width=1\linewidth]{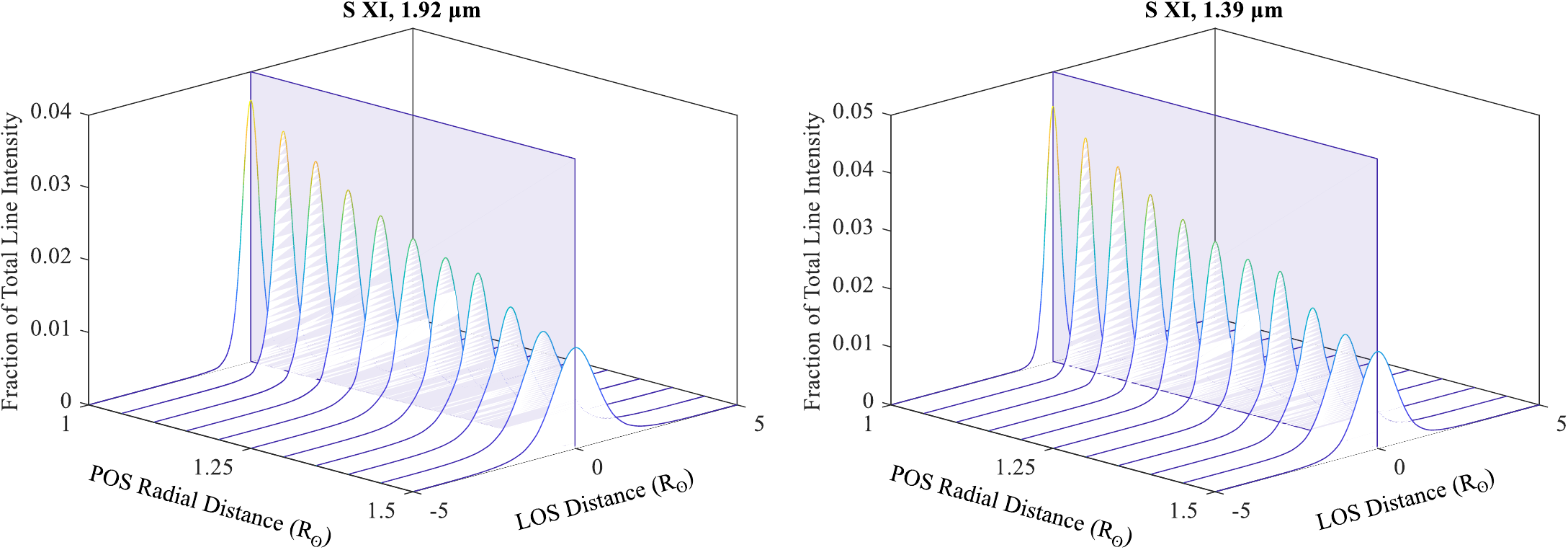}
	\caption{Fraction of line intensity contributed by each 0.025 \Rs\ 
          along the line of sight. Each curve sums to one along the ‘LOS Distance’ coordinate. The 1.92~\mic\ line has a smaller contribution from the POS than the 1.39~\mic\ line, resulting in a lower line ratio (and higher density) when only the POS intensity is taken into account instead of the entire LOS integration. }
	\label{fig:s_11_emiss}
\end{figure}

In order to estimate the radial (LOS-corrected) density profile, we use the spherically symmetric model described in \citet{delzanna_etal:2018_cosie} to remove the AIR-Spec intensity contribution from outside the plane of the sky (POS).  Figure \ref{fig:s_11_emiss} shows the fraction of the total line intensity contributed by each \textbf{0.025 \Rs} along the LOS, as predicted by the model. Only 4--5\% of the contribution comes from the  \textbf{0.025 \Rs}  nearest the POS. If we scale the AIR-Spec \ion{S}{11} intensities by this fraction before taking the line ratio, we obtain the radial density estimates in the right half of Figure \ref{fig:density1}.

\begin{figure}[!htbp]
	\centering
	\includegraphics[angle=0,width=0.35\linewidth]{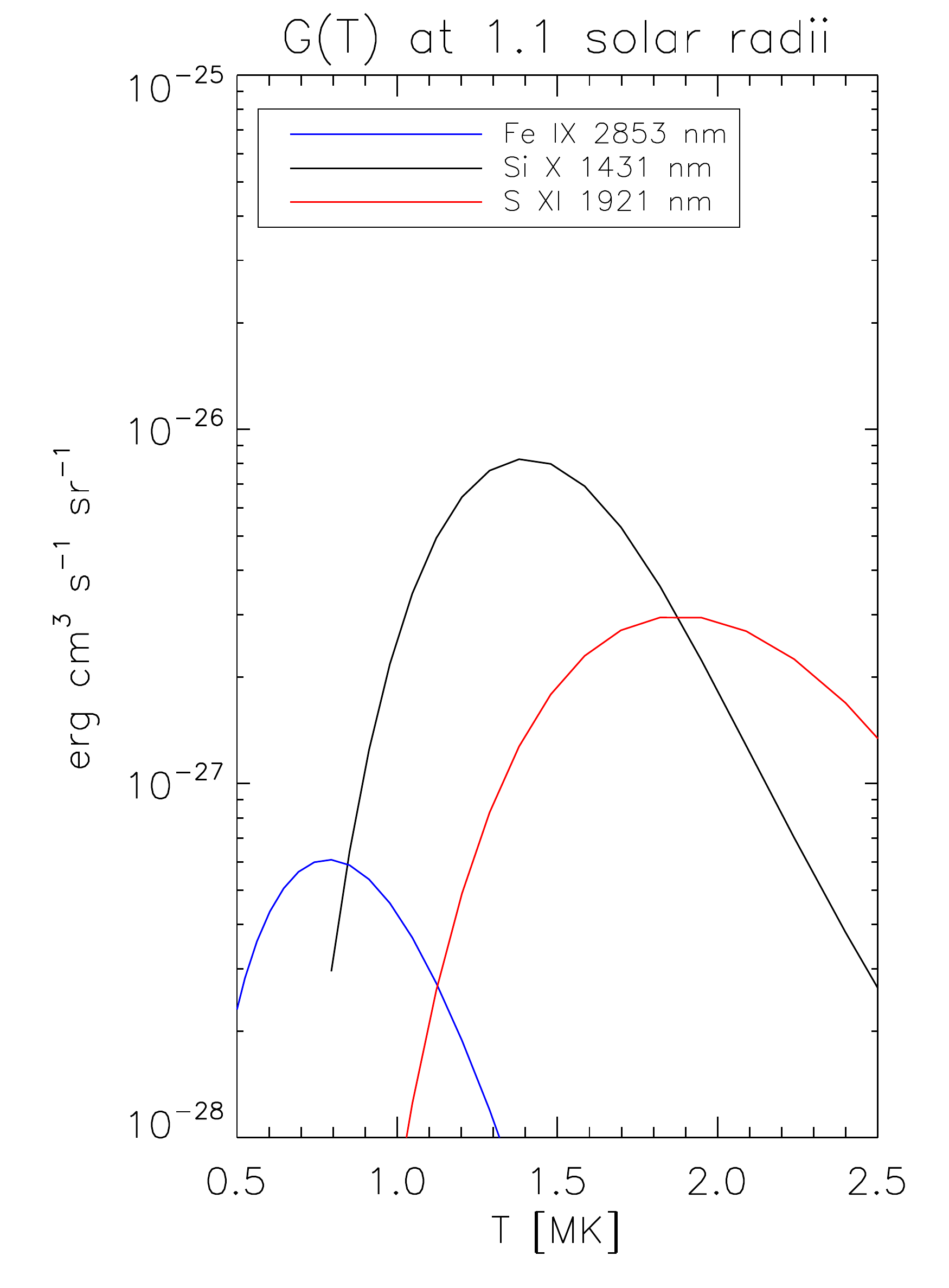}
	\caption{Contribution functions of the main AIR-Spec lines, calculated 
	with photoexcitation at 1.1\rsun. }
	\label{fig:gt}
\end{figure}

{Figure~\ref{fig:gt} shows the emissivities (contribution functions, calculated in ionization equilibrium) of the main AIR-Spec lines, indicating the range of temperatures where the lines can be formed. We use the emissivity ratio of the two strongest lines, \ion{Si}{10} 1.43 \mic\ and \ion{S}{11} 1.92 \mic, to estimate the temperatures shown in Figures \ref{fig:temps_by_instrument}c and \ref{fig:temps_by_region}. The contribution functions are computed using photoexcitation and integrated along the LOS before the ratio is taken. Because the modeled line ratio includes emission from the entire LOS, the LOS-averaged AIR-Spec radiances are used to map line ratio to temperature.}

\begin{figure}[!htbp]
	\centering
	\includegraphics[angle=0,width=0.7\linewidth]{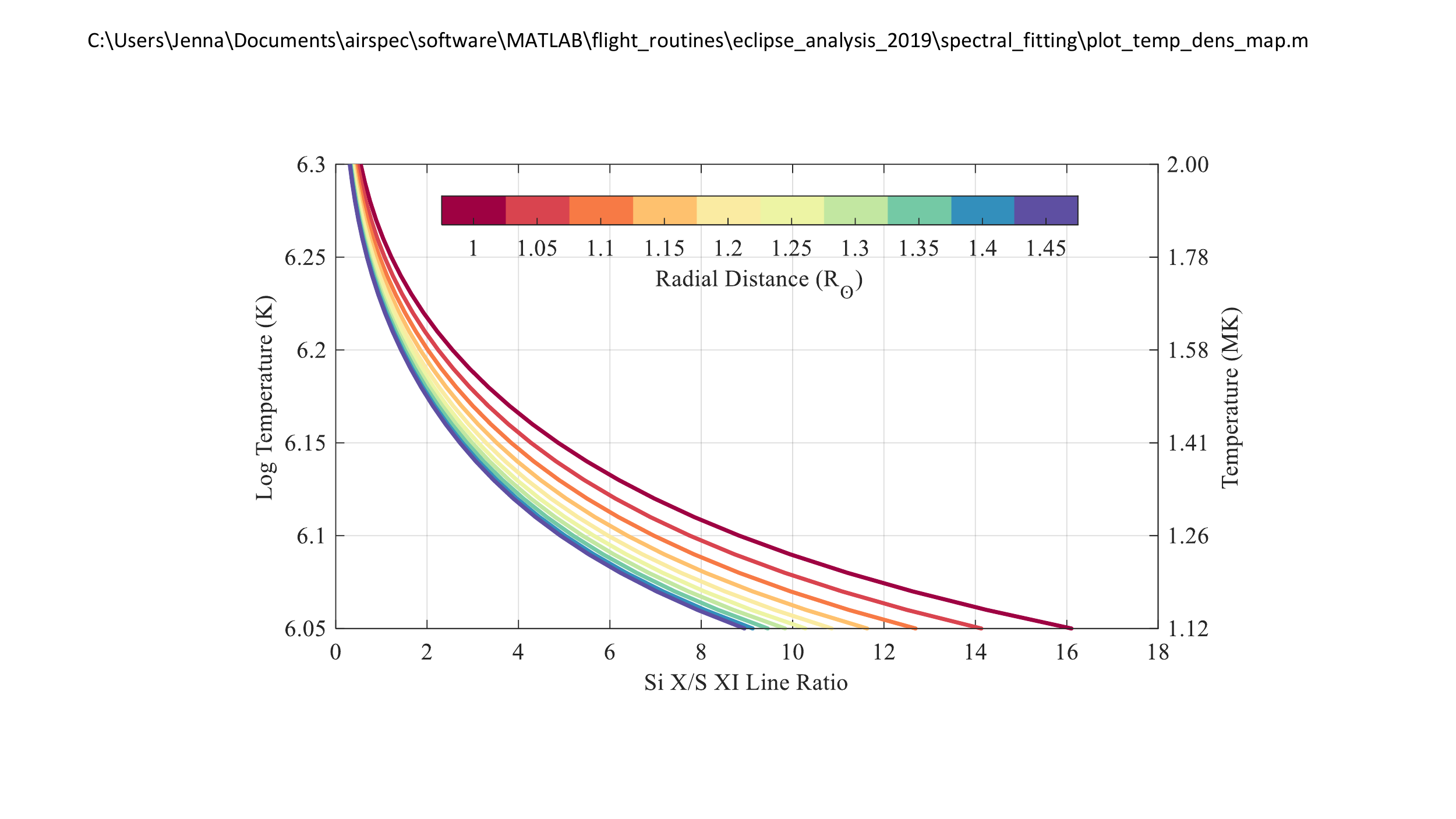}
	\caption{Temperature as a function of \ion{Si}{10}/\ion{S}{11} 1.43/1.92 \mic\ line ratio and distance from Sun center.}
	\label{fig:temp_mapping}
\end{figure}

{ As with the density mapping, the distance from the underlying photospheric radiation field is an important parameter in the mapping from line ratio to temperature.  Figure \ref{fig:temp_mapping} shows the dependence of temperature on the combination of the \ion{Si}{10} 1.43 \mic/\ion{S}{11} 1.92 \mic\ line ratio and the distance from Sun center. The line ratio is sensitive to temperatures in the 1--2 MK range.}

 \section{Temperature distribution from EIS}
\label{sec:eis_temp}

 \begin{figure}[!htbp]
	\centering
	\includegraphics[angle=0,width=0.6\linewidth]{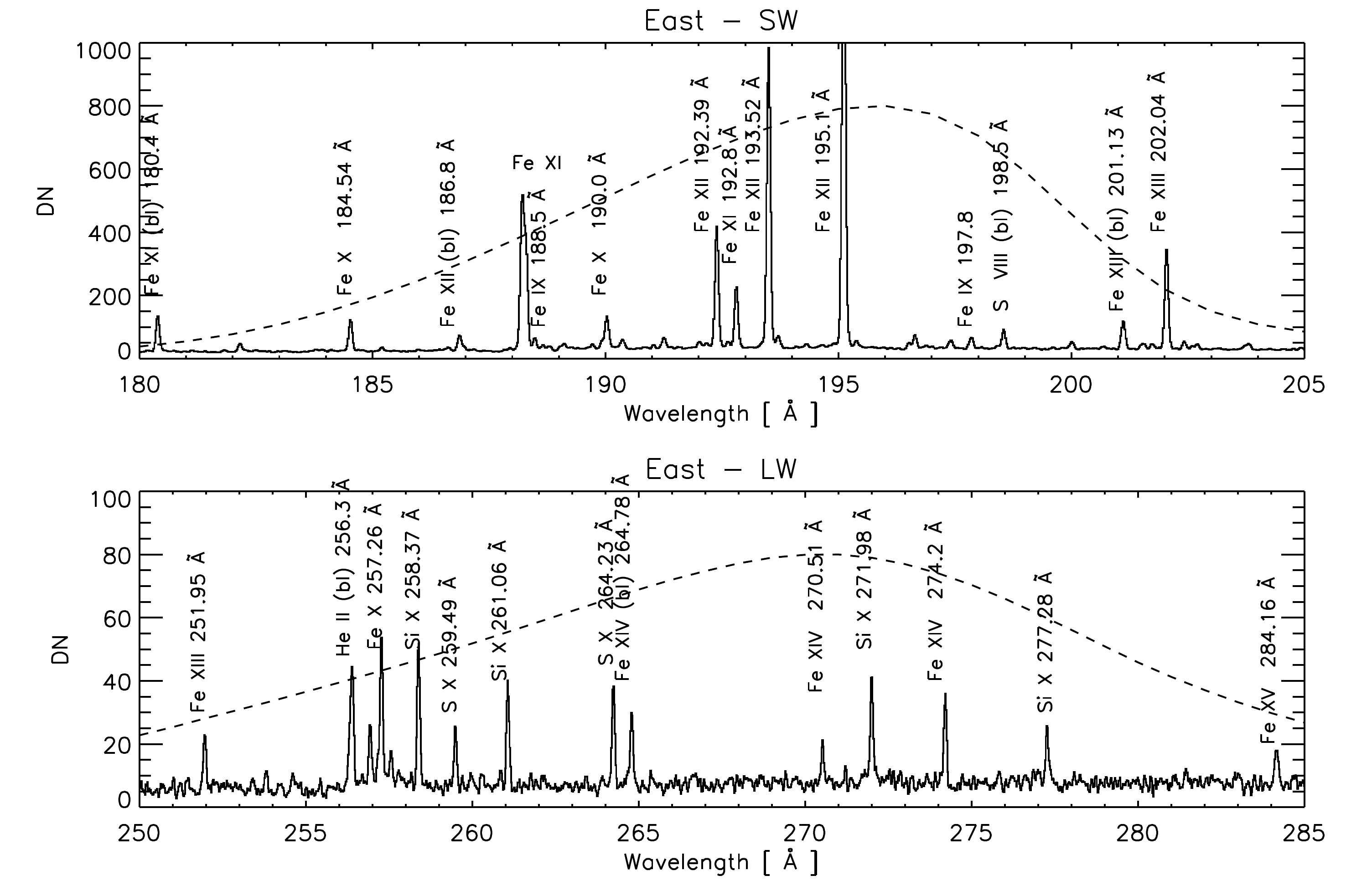}
	\includegraphics[angle=0,width=0.6\linewidth]{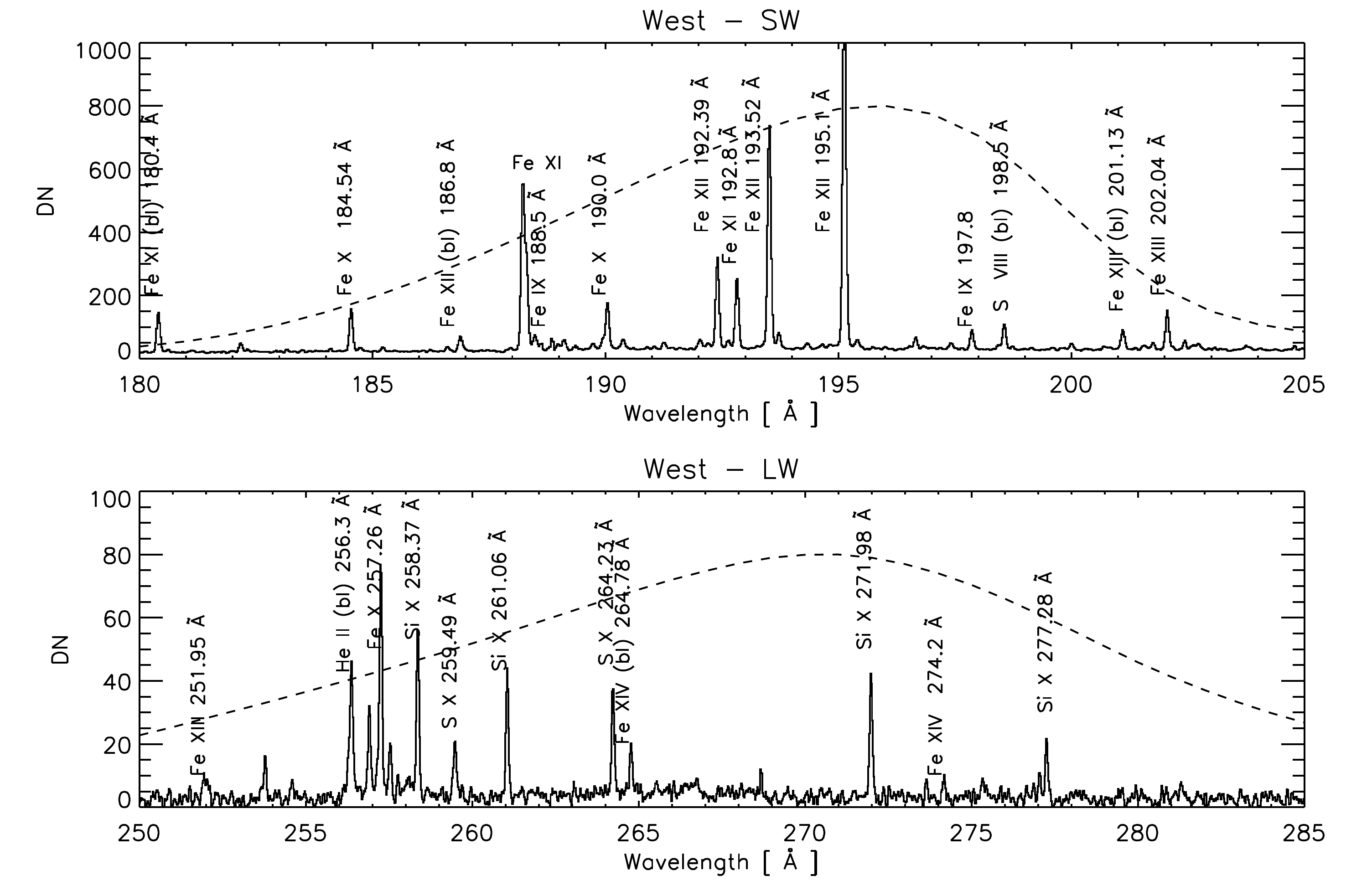}
	\caption{Hinode EIS spectra for the  off-limb
          observation above the east and west limb (120 s, 2\arcsec\ slit), around 1.09~\rsun. The dashed lines
       are the scaled ground-calibration effective area, to show why spectral lines in
        the middle of the two SW and LW detectors are stronger.}
	\label{fig:eis_sp_airspec}
\end{figure}

 Figure~\ref{fig:eis_sp_airspec} shows the averaged EIS spectra
 in the east and west regions, used to obtain the DEM distribution. 
 The strongest lines are from \ion{Fe}{10}--\ion{Fe}{13}, and
 \ion{Si}{10}. The west region has significant emission in 
 \ion{Fe}{14} and \ion{Fe}{15} which is not present on the east. 
The \ion{S}{10} lines are well observed and provide a strong constraint
on the relative FIP bias between high-FIP and low-FIP elements.
 
 Table~\ref{tab:lines_east} shows the results of the DEM analysis on the
off-limb observation above the east limb. Only a few of the lines are shown.
$T_{\rm max}$ is the temperature of the maximum of the
contribution function $G(T)$, while  the effective temperature $T_{\rm eff}$ is:
\beq
T_{\rm eff} = \int G{\left({T}\right)}~
DEM{\left({T}\right)} ~T~dT / 
{\int G{\left({T}\right)}~DEM{\left({T}\right)}~dT} \quad .
\eeq

 For each observed line, the CHIANTI
programs search for all the transitions (that form a blend) within
a wavelength range, centred on the observed wavelength.
Only the main lines contributing to the blend are listed, with their fractional
contribution indicated in the last column.
 Table~\ref{tab:lines_west} shows the results of the DEM analysis on the
off-limb observation above the west limb.  Only a few of the lines are shown.

\begin{table}
  \caption{Observed  and predicted radiances for the  off-limb
    observation above the east limb (120 s, 2\arcsec\ slit). 
    $\lambda_{\rm obs}$ (\AA) are the observed wavelengths,
DN is the number of total counts in each line, while $I_{\rm obs}$
is the calibrated radiance (erg cm$^{-2}$ s$^{-1}$ sr$^{-1}$) obtained with our new EIS calibration.
$T_{\rm max}$ and $T_{\rm eff}$ are the maximum and effective temperature 
(log values, in K;  see text),
$R$ the ratio between the predicted and observed radiances, 
Ion and $\lambda_{\rm exp}$ (\AA) the main contributing ion, and 
$r$ the fractional contribution to the blend. } 
\begin{center}
\begin{tabular}{@{}rrrccrlrr@{}}
\hline\hline \noalign{\smallskip}
 $\lambda_{\rm obs}$  & DN & $I_{\rm obs}$   & log $T_{\rm max}$ & log $T_{\rm eff}$  & $R$ & Ion & $\lambda_{\rm exp}$   &  $r$ \\
\noalign{\smallskip}\hline\noalign{\smallskip}

 185.20 &       47 & 5.6 &  5.73 &  6.06 &  0.83 &  \ion{Fe}{8} &  185.213 & 0.94 \\ 
 
 186.61 &       57 & 4.7 &  5.72 &  6.06 &  0.84 &  \ion{Fe}{8} &  186.598 & 0.81 \\

 189.93 &       80 & 3.5 &  5.92 &  6.10 &  1.10 &  \ion{Fe}{9} &  189.935 & 0.95 \\ 
 
 188.48 &      142 & 8.9 &  5.93 &  6.10 &  1.06 &  \ion{Mn}{9} &  188.480 & 0.11 \\ 
                               &  &   &  &  &  &  \ion{Fe}{9} &  188.493 & 0.83 \\ 
 
 197.85 &      128 & 3.8 &  5.93 &  6.10 &  1.15 &  \ion{Fe}{9} &  197.854 & 0.93 \\ 
 
 257.27 &      213 & 41.6 &  6.02 &  6.11 &  0.98 &  \ion{Fe}{10} &  257.259 & 0.21 \\ 
                               &  &   &  &  &  &  \ion{Fe}{10} &  257.261 & 0.77 \\ 
 
 184.53 &      403 & 56.4 &  6.03 &  6.12 &  0.80 &  \ion{Fe}{10} &  184.537 & 0.96 \\ 
 
 190.03 &      391 & 16.5 &  6.03 &  6.12 &  1.00 &  \ion{Fe}{10} &  190.037 & 0.86 \\

 198.54 &      236 & 7.2 &  5.95 &  6.12 &  1.45 &  \ion{S}{8} &  198.553 & 0.44 \\ 
                              &   &   &  &  &  &  \ion{Fe}{11} &  198.538 & 0.51 \\ 
 
 192.81 &      811 & 26.3 &  5.42 &  6.13 &  1.15 &  \ion{Fe}{11} &  192.813 & 0.97 \\

 188.21 &     1940 & 131.0 &  6.12 &  6.14 &  1.10 &  \ion{Fe}{11} &  188.216 & 0.98 \\ 
 
 188.29 &     1370 & 90.2 &  6.12 &  6.14 &  0.98 &  \ion{Fe}{11} &  188.299 & 0.98 \\ 
 
 180.39 &      475 & 337.0 &  6.12 &  6.14 &  0.90 &  \ion{Fe}{11} &  180.401 & 0.98 \\ 
 
 258.38 &      201 & 37.7 &  6.15 &  6.14 &  1.03 &  \ion{Si}{10} &  258.374 & 0.97 \\ 
 
 272.00 &      151 & 20.1 &  6.15 &  6.14 &  1.00 &  \ion{Si}{10} &  271.992 & 0.98 \\ 
 
 261.07 &      137 & 21.9 &  6.15 &  6.14 &  1.12 &  \ion{Si}{10} &  261.056 & 0.98 \\ 
 
 277.27 &       73 & 14.9 &  6.15 &  6.14 &  1.04 &  \ion{Si}{10} &  277.264 & 0.98 \\ 
 
 259.49 &       80 & 14.4 &  6.18 &  6.15 &  1.01 &  \ion{S}{10} &  259.496 & 0.98 \\ 
 
 264.24 &      133 & 18.7 &  6.18 &  6.15 &  1.13 &  \ion{S}{10} &  264.230 & 0.98 \\

 193.50 &     3860 & 117.0 &  6.19 &  6.15 &  1.10 &  \ion{Fe}{12} &  193.509 & 0.94 \\ 
 
 195.11 &     6460 & 169.0 &  6.19 &  6.15 &  1.09 &  \ion{Fe}{12} &  195.119 & 0.97 \\ 
 
 192.39 &     1580 & 54.7 &  6.19 &  6.15 &  1.08 &  \ion{Fe}{12} &  192.394 & 0.97 \\ 
  
 200.01 &      113 & 4.1 &  6.24 &  6.17 &  1.14 &  \ion{Fe}{13} &  200.021 & 0.91 \\ 
 
 191.25 &      124 & 4.8 &  6.28 &  6.17 &  1.03 &  \ion{S}{11} &  191.266 & 0.94 \\ 
 
 251.97 &       67 & 21.3 &  6.24 &  6.17 &  1.05 &  \ion{Fe}{13} &  251.952 & 0.97 \\ 
 
 202.04 &     1260 & 113.0 &  6.24 &  6.17 &  0.86 &  \ion{Fe}{13} &  202.044 & 0.97 \\ 
 
 264.79 &       94 & 13.2 &  6.29 &  6.17 &  1.20 &  \ion{Fe}{14} &  264.788 & 0.61 \\ 
                               &  &   &  &  &  &  \ion{Fe}{11} &  264.772 & 0.36 \\ 
 
 274.21 &      125 & 17.9 &  6.29 &  6.19 &  0.86 &  \ion{Fe}{14} &  274.203 & 0.95 \\

 270.52 &       57 & 8.1 &  6.29 &  6.19 &  0.82 &  \ion{Fe}{14} &  270.520 & 0.97 \\ 
 
 262.98 &        4 & 0.61 &  6.43 &  6.20 &  1.16 &  \ion{Fe}{16} &  262.976 & 0.26 \\ 
                              &   &   &  &  &  &  \ion{Fe}{13} &  262.984 & 0.66 \\ 
 
 284.16 &       47 & 20.5 &  6.34 &  6.22 &  1.05 &  \ion{Fe}{15} &  284.163 & 0.96 \\

 \noalign{\smallskip}\hline 
\end{tabular}
\end{center}
\normalsize
 \label{tab:lines_east} 
\end{table}

\begin{table}[!htbp]
  \caption{Observed  and predicted radiances for the  off-limb
    observation above the west limb (120 s, 2\arcsec\ slit). 
    The notation is the same as  Table~\ref{tab:lines_east}.
    } 
\begin{center}
\begin{tabular}{@{}rrrccrlrr@{}}
\hline\hline \noalign{\smallskip}
 $\lambda_{\rm obs}$  & DN & $I_{\rm obs}$   & log $T_{\rm max}$ & log $T_{\rm eff}$  & $R$ & Ion & $\lambda_{\rm exp}$   &  $r$ \\
\noalign{\smallskip}\hline\noalign{\smallskip}
 
 185.22 &       48 & 5.8 &  5.73 &  6.07 &  1.08 &  \ion{Fe}{8} &  185.213 & 0.97 \\

 189.94 &      109 & 4.7 &  5.92 &  6.08 &  1.08 &  \ion{Fe}{9} &  189.935 & 0.98 \\ 
 
 197.86 &      231 & 6.9 &  5.93 &  6.08 &  0.92 &  \ion{Fe}{9} &  197.854 & 0.95 \\ 
 
 257.25 &      312 & 61.1 &  6.02 &  6.09 &  0.88 &  \ion{Fe}{10} &  257.259 & 0.20 \\ 
                               &  &   &  &  &  &  \ion{Fe}{10} &  257.261 & 0.78 \\ 
 
 184.54 &      538 & 75.3 &  6.03 &  6.09 &  0.75 &  \ion{Fe}{10} &  184.537 & 0.97 \\ 
 
 190.04 &      575 & 24.3 &  6.03 &  6.09 &  0.83 &  \ion{Fe}{10} &  190.037 & 0.90 \\ 
 
 198.56 &      293 & 9.0 &  5.95 &  6.09 &  1.26 &  \ion{S}{8} &  198.553 & 0.54 \\ 
                              &   &   &  &  &  &  \ion{Fe}{11} &  198.538 & 0.42 \\

 256.91 &      117 & 23.2 &  6.10 &  6.09 &  0.96 &  \ion{Fe}{11} &  256.919 & 0.98 \\ 
 
 180.41 &      513 & 364.0 &  6.11 &  6.10 &  0.83 &  \ion{Fe}{11} &  180.401 & 0.93 \\

 188.22 &     2040 & 137.0 &  6.12 &  6.10 &  0.99 &  \ion{Fe}{11} &  188.216 & 0.98 \\ 
 
 192.82 &      877 & 28.4 &  6.12 &  6.10 &  1.01 &  \ion{Fe}{11} &  192.813 & 0.97 \\ 
 
 256.37 &      223 & 46.1 &  6.16 &  6.10 &  0.81 &  \ion{Si}{10} &  256.377 & 0.65 \\ 
                              &   &   &  &  &  &  \ion{Fe}{12} &  256.410 & 0.14 \\ 
                              &   &   &  &  &  &  \ion{Fe}{10} &  256.398 & 0.12 \\ 
 
 258.36 &      219 & 41.2 &  6.15 &  6.10 &  0.72 &  \ion{Si}{10} &  258.374 & 0.97 \\ 
 
 271.98 &      173 & 23.0 &  6.15 &  6.10 &  0.73 &  \ion{Si}{10} &  271.992 & 0.98 \\ 
 
 261.05 &      166 & 26.6 &  6.15 &  6.10 &  0.76 &  \ion{Si}{10} &  261.056 & 0.98 \\

 264.22 &      140 & 19.7 &  6.18 &  6.10 &  0.80 &  \ion{S}{10} &  264.230 & 0.98 \\

 193.52 &     2850 & 86.2 &  6.19 &  6.10 &  0.99 &  \ion{Fe}{12} &  193.509 & 0.94 \\ 
 
 192.40 &     1100 & 38.1 &  6.19 &  6.10 &  1.02 &  \ion{Fe}{12} &  192.394 & 0.97 \\ 
 
 195.13 &     4660 & 122.0 &  6.19 &  6.10 &  0.99 &  \ion{Fe}{12} &  195.119 & 0.97 \\ 
 
 264.77 &       63 & 8.8 &  6.29 &  6.11 &  0.88 &  \ion{Fe}{14} &  264.788 & 0.26 \\ 
                              &   &   &  &  &  &  \ion{Fe}{11} &  264.772 & 0.71 \\ 
  
 200.00 &       62 & 2.3 &  6.24 &  6.11 &  0.70 &  \ion{Fe}{13} &  200.021 & 0.81 \\ 
 
 191.26 &       61 & 2.4 &  6.28 &  6.11 &  0.95 &  \ion{S}{11} &  191.266 & 0.91 \\

 202.05 &      471 & 42.2 &  6.24 &  6.12 &  0.96 &  \ion{Fe}{13} &  202.044 & 0.97 \\ 
 
 274.17 &       22 & 3.3 &  6.29 &  6.13 &  1.09 &  \ion{Fe}{14} &  274.203 & 0.89 \\ 
                                  &   &  &  &  &  \ion{Fe}{11} &  256.654 & 0.18 \\ 

 \noalign{\smallskip}\hline 
\end{tabular}
\end{center}
\normalsize
 \label{tab:lines_west} 
\end{table}
    
\section{EIS calibration}
\label{sec:eis_cal}

We performed a DEM analysis of
a quiet Sun off-limb observations close in time to the eclipse, on 2019-
06-28 at 09:38 UT, with the 2\arcsec\ slit and 60 s exposures.
Table~\ref{tab:list3} gives a list of the main lines. 
The emissivities were calculated at a constant density log Ne =8.6.
The EIS effective areas were adjusted to match observation. 
This calibration was then used for the Eclipse observations.
It is clear that 
all the main lines are well represented within $\pm$15\%.

\begin{table}
  \caption{Observed  and predicted radiances for the  off-limb
    observation (60s, 2\arcsec\ slit) on 2019-Jun-28 used for the EIS radiometric calibration.
    The notation is the same as  Table~\ref{tab:lines_east}.
  }
\begin{center} 
\footnotesize
\begin{tabular}{@{}lllllllll@{}}
 \hline\hline \noalign{\smallskip}
  $\lambda_{\rm obs}$  & DN & $I_{\rm obs}$   &  $T_{\rm max}$ & $T_{\rm eff}$  & $R$ & Ion & $\lambda_{\rm exp}$   &  $r$ \\
 \hline \noalign{\smallskip}

 194.66 &      417 & 22.6 &  5.72 &  6.05 &  1.04 &  \ion{Fe}{8} &  194.661 & 0.95 \\ 
 
 185.22 &      412 & 97.0 &  5.73 &  6.05 &  1.17 &  \ion{Fe}{8} &  185.213 & 0.96 \\ 
 
 186.60 &      464 & 75.6 &  5.73 &  6.06 &  1.17 &  \ion{Fe}{8} &  186.598 & 0.85 \\ 
 
 
 189.94 &      785 & 67.4 &  5.94 &  6.08 &  1.23 &  \ion{Fe}{9} &  189.935 & 0.97 \\ 
 
 188.49 &     1290 & 162.0 &  5.95 &  6.08 &  1.00 &  \ion{Fe}{9} &  188.493 & 0.89 \\ 
 
 197.85 &     1320 & 78.9 &  5.96 &  6.09 &  1.04 &  \ion{Fe}{9} &  197.854 & 0.94 \\ 
 
 257.26 &     1510 & 589.0 &  6.04 &  6.10 &  0.95 &  \ion{Fe}{10} &  257.259 & 0.25 \\ 
                              &   &   &  &  &  &  \ion{Fe}{10} &  257.263 & 0.73 \\ 
 
 177.24 &      346 & 1650.0 &  6.05 &  6.11 &  1.08 &  \ion{Fe}{10} &  177.240 & 0.98 \\ 
 
 174.53 &      177 & 2660.0 &  6.05 &  6.11 &  1.19 &  \ion{Fe}{10} &  174.531 & 0.98 \\ 
 
 184.54 &     2280 & 637.0 &  6.05 &  6.11 &  1.06 &  \ion{Fe}{10} &  184.537 & 0.96 \\ 
 
 190.04 &     3050 & 257.0 &  6.06 &  6.12 &  1.09 &  \ion{Fe}{12} &  190.040 & 0.12 \\ 
                              &   &   &  &  &  &  \ion{Fe}{10} &  190.037 & 0.76 \\ 
 
 
 184.79 &      164 & 43.3 &  6.12 &  6.13 &  1.20 &  \ion{Fe}{11} &  184.793 & 0.93 \\ 
 
 257.55 &      441 & 170.0 &  6.12 &  6.13 &  1.07 &  \ion{Fe}{11} &  257.554 & 0.70 \\ 
                             &    &   &  &  &  &  \ion{Fe}{11} &  257.547 & 0.25 \\ 
 
 189.02 &      427 & 47.3 &  6.13 &  6.13 &  0.87 &  \ion{Fe}{11} &  188.997 & 0.95 \\ 
 
 256.92 &      608 & 241.0 &  6.12 &  6.13 &  0.86 &  \ion{Fe}{11} &  256.919 & 0.92 \\ 
 
 
 181.13 &      150 & 167.0 &  6.13 &  6.13 &  0.89 &  \ion{Fe}{11} &  181.130 & 0.97 \\ 
 
 192.81 &     5040 & 326.0 &  5.42 &  6.13 &  0.91 &  \ion{Fe}{11} &  192.813 & 0.96 \\ 
 
 188.22 &     9190 & 1240.0 &  6.13 &  6.13 &  1.13 &  \ion{Fe}{11} &  188.216 & 0.98 \\ 
 
 188.30 &     6090 & 801.0 &  6.13 &  6.13 &  1.06 &  \ion{Fe}{11} &  188.299 & 0.98 \\ 
 
 
 180.40 &     1820 & 2570.0 &  6.13 &  6.13 &  1.13 &  \ion{Fe}{11} &  180.401 & 0.98 \\ 
 
 202.71 &      424 & 87.6 &  6.13 &  6.14 &  1.37 &  \ion{Fe}{11} &  202.705 & 0.93 \\ 
 
 258.37 &     1460 & 546.0 &  6.15 &  6.14 &  0.96 &  \ion{Si}{10} &  258.374 & 0.97 \\ 
 
 253.79 &      223 & 115.0 &  6.15 &  6.14 &  0.90 &  \ion{Si}{10} &  253.790 & 0.97 \\ 
 
 271.99 &      852 & 226.0 &  6.15 &  6.14 &  0.93 &  \ion{Si}{10} &  271.992 & 0.97 \\ 
 
 261.06 &      853 & 273.0 &  6.15 &  6.14 &  0.92 &  \ion{Si}{10} &  261.056 & 0.98 \\ 
 
 277.27 &      452 & 181.0 &  6.15 &  6.14 &  0.89 &  \ion{Si}{10} &  277.264 & 0.98 \\ 
 
 
 
 249.39 &      107 & 89.1 &  6.19 &  6.16 &  1.12 &  \ion{Fe}{12} &  249.388 & 0.69 \\ 
                              &   &   &  &  &  &  \ion{Fe}{12} &  249.384 & 0.22 \\ 
 
 
 
 
 
 193.51 &    18200 & 1100.0 &  6.19 &  6.17 &  1.13 &  \ion{Fe}{12} &  193.509 & 0.94 \\ 
 
 192.39 &     8090 & 559.0 &  6.19 &  6.17 &  1.02 &  \ion{Fe}{12} &  192.394 & 0.96 \\ 
 
 195.12 &    27800 & 1450.0 &  6.19 &  6.17 &  1.23 &  \ion{Fe}{12} &  195.119 & 0.97 \\ 
 
 
 200.02 &     1510 & 109.0 &  6.25 &  6.20 &  0.90 &  \ion{Fe}{13} &  200.021 & 0.94 \\ 
 
 
 246.21 &      170 & 185.0 &  6.25 &  6.20 &  0.80 &  \ion{Fe}{13} &  246.209 & 0.95 \\ 
 
 
 251.95 &      421 & 268.0 &  6.25 &  6.20 &  1.02 &  \ion{Fe}{13} &  251.952 & 0.96 \\ 
 
 204.94 &      180 & 59.2 &  6.25 &  6.20 &  1.18 &  \ion{Fe}{13} &  204.942 & 0.96 \\ 
 
 209.92 &      135 & 149.0 &  6.25 &  6.20 &  1.05 &  \ion{Fe}{13} &  209.916 & 0.95 \\ 
 
 202.04 &     5070 & 906.0 &  6.25 &  6.20 &  0.98 &  \ion{Fe}{13} &  202.044 & 0.96 \\ 
 
 264.79 &      858 & 239.0 &  6.29 &  6.20 &  0.99 &  \ion{Fe}{14} &  264.789 & 0.72 \\ 
                              &   &   &  &  &  &  \ion{Fe}{11} &  264.772 & 0.24 \\ 
 
 274.20 &      835 & 240.0 &  6.29 &  6.22 &  1.05 &  \ion{Fe}{14} &  274.204 & 0.94 \\ 
 
 252.20 &       58 & 36.3 &  6.29 &  6.22 &  1.20 &  \ion{Fe}{14} &  252.200 & 0.94 \\ 
 
 
 211.32 &      370 & 418.0 &  6.29 &  6.22 &  1.06 &  \ion{Fe}{14} &  211.317 & 0.96 \\ 
 
 257.40 &      180 & 70.1 &  6.29 &  6.22 &  1.14 &  \ion{Fe}{14} &  257.394 & 0.96 \\ 
 
 270.52 &      398 & 112.0 &  6.29 &  6.22 &  0.99 &  \ion{Fe}{14} &  270.521 & 0.96 \\ 
 
 
 284.16 &      445 & 386.0 &  6.34 &  6.24 &  1.23 &  \ion{Fe}{15} &  284.163 & 0.96 \\ 

\noalign{\smallskip}\hline                                   
\end{tabular}
\normalsize
\end{center}
\label{tab:list3}
\end{table}

\begin{figure}[!htbp]
	\centerline{
	\includegraphics[angle=-90,width=0.5\linewidth]{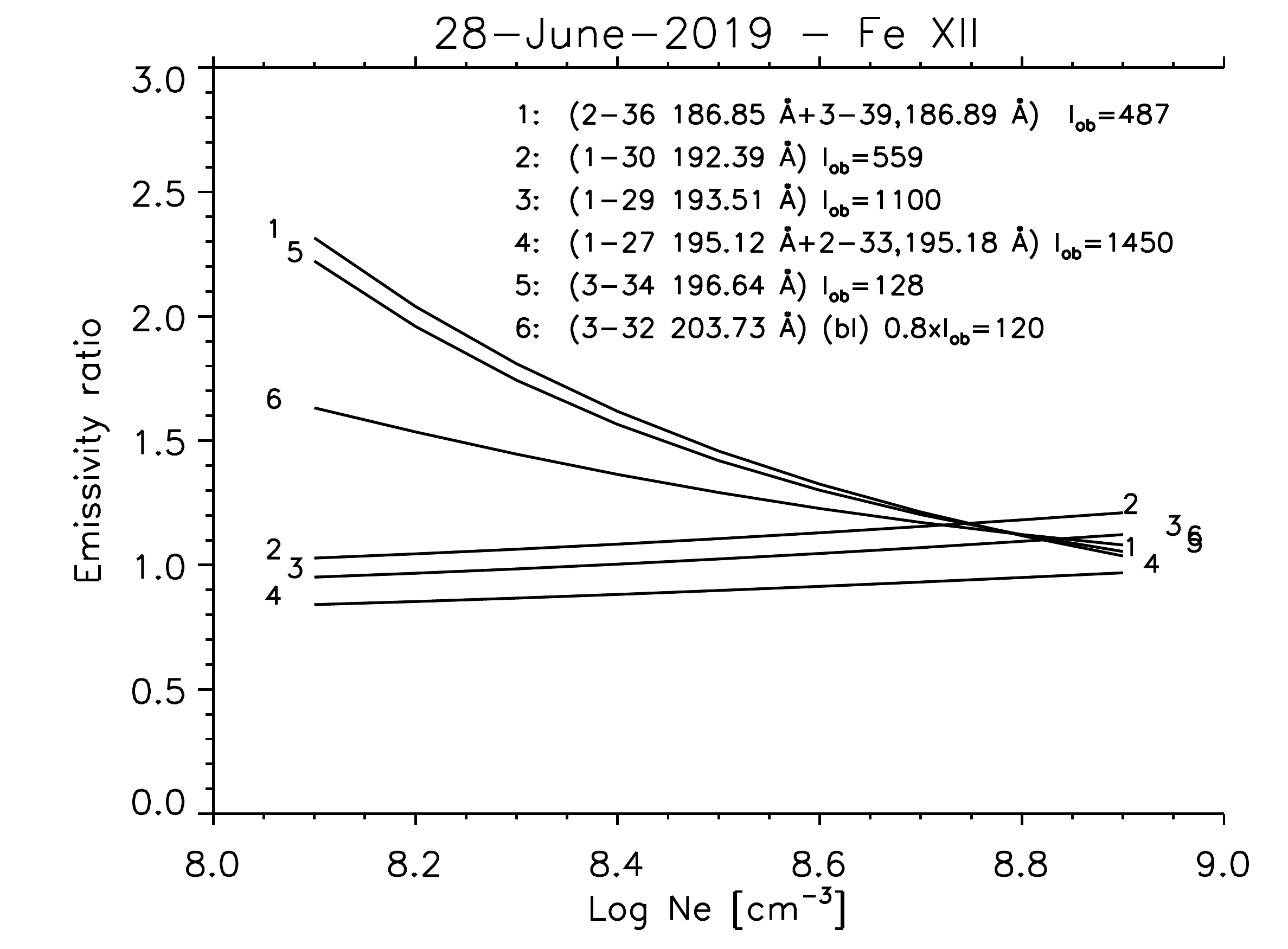}
		\includegraphics[angle=-90,width=0.5\linewidth]{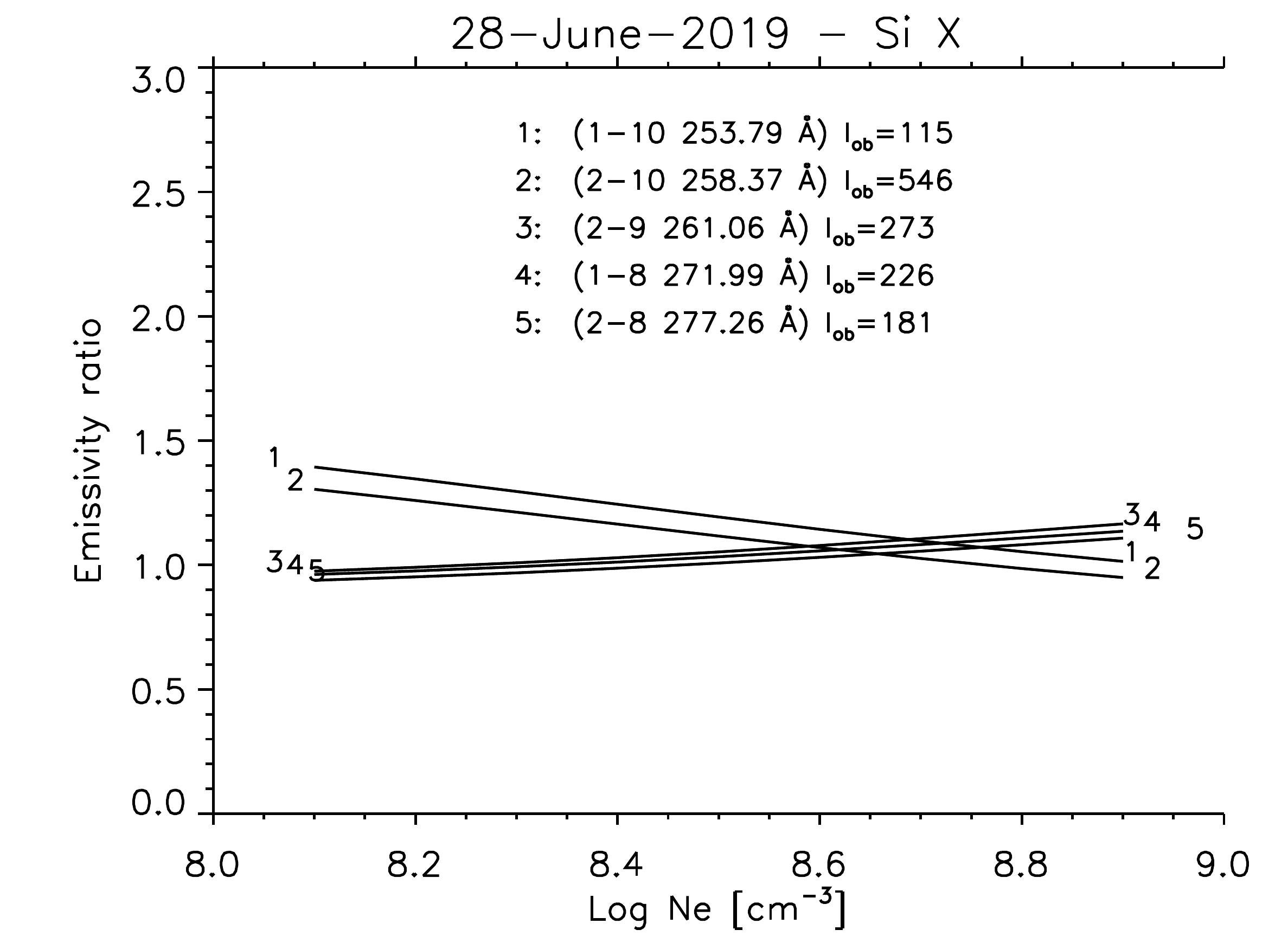}}
	\centerline{
	\includegraphics[angle=-90,width=0.5\linewidth]{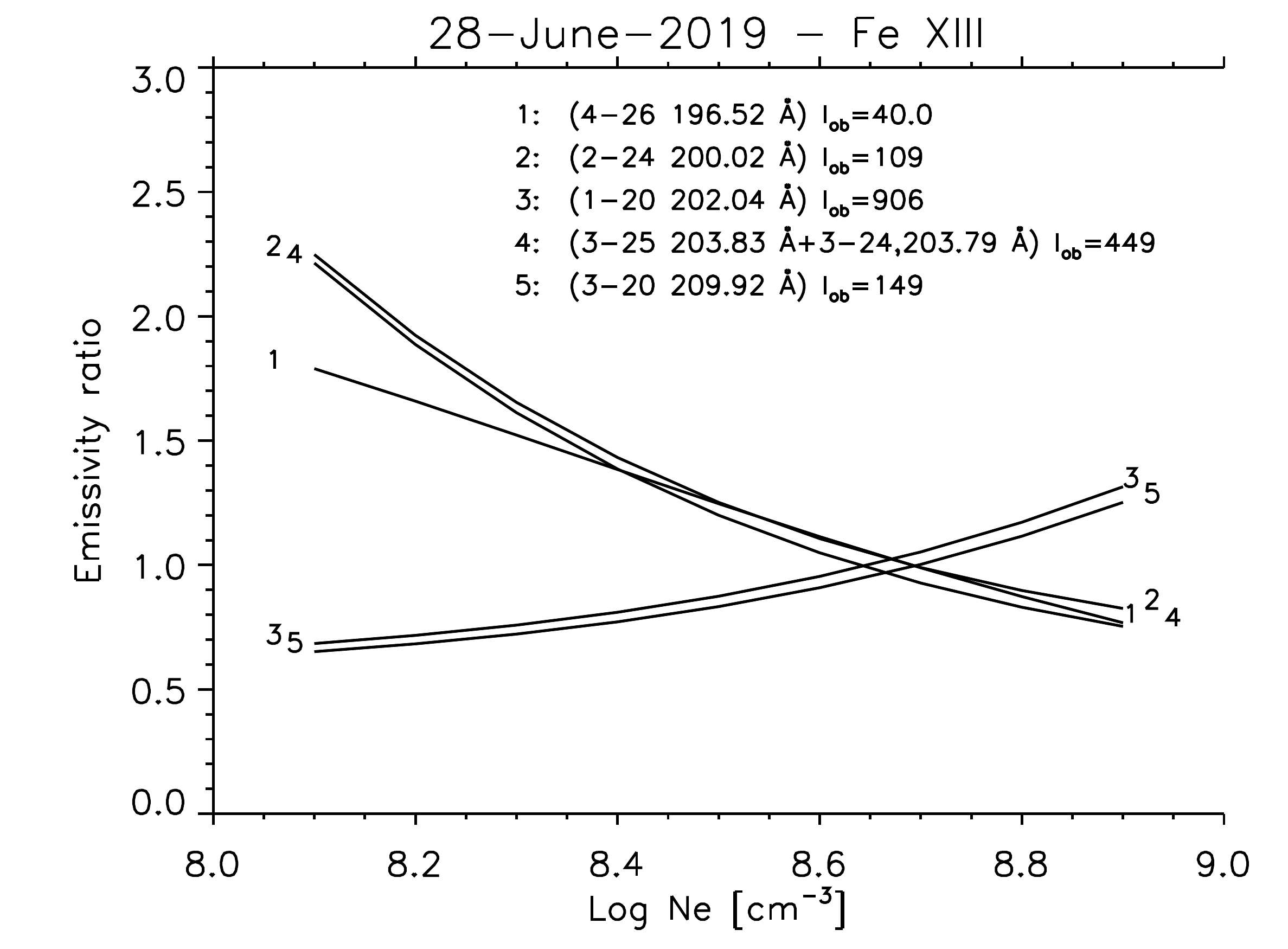}
		\includegraphics[angle=-90,width=0.5\linewidth]{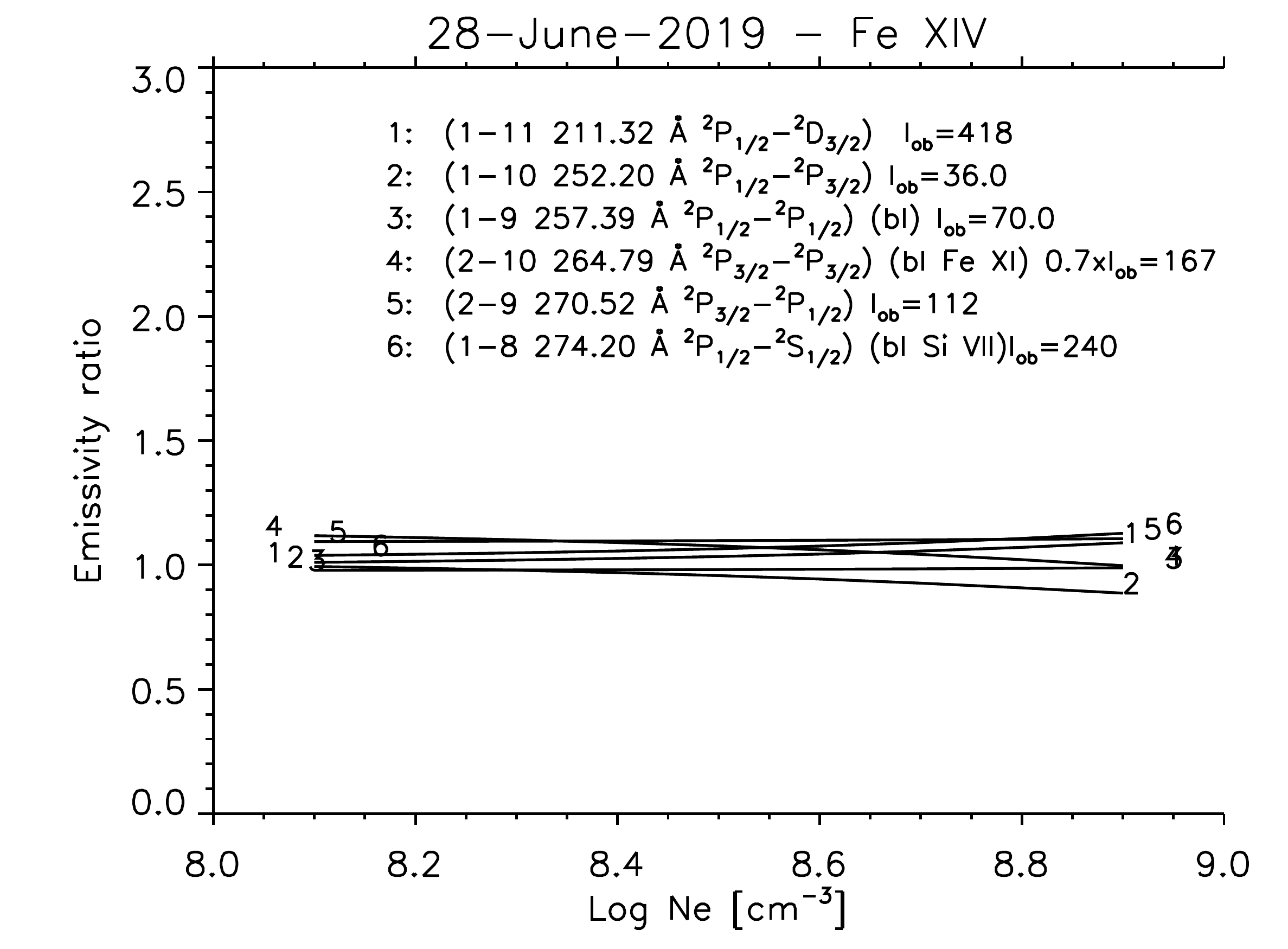}
		}
	\caption{Emissivity ratios of a selection of EIS lines for the off-limb quiet Sun observation of 2019-06-28.}
	\label{fig:eis_cal_emissivities}
\end{figure}

Figure~\ref{fig:eis_cal_emissivities} shows the emissivity ratios of  a selection of lines,
from four of the main ions observed by EIS.
The emissivity ratios are essentially the ratios of the observed intensities 
vs. the theoretical ones, as a function of density, calculated for a 
constant temperature \citep[see][]{delzanna_etal:2004}.
It is clear that the main density-diagnostic ratios from Fe XII, Fe XIII, and Si X give
consistent densities. 
Note that the main Fe XII lines, at 192,193,195~\AA\ are somewhat 
affected by opacity (i.e. their observed intensities are lower than expected),
a typical feature of active region and off-limb observations, as 
shown for the first time by \cite{delzanna_etal:2019_ntw}.

\begin{figure}[!htbp]
	\centering
	\includegraphics[angle=0,width=0.9\linewidth]{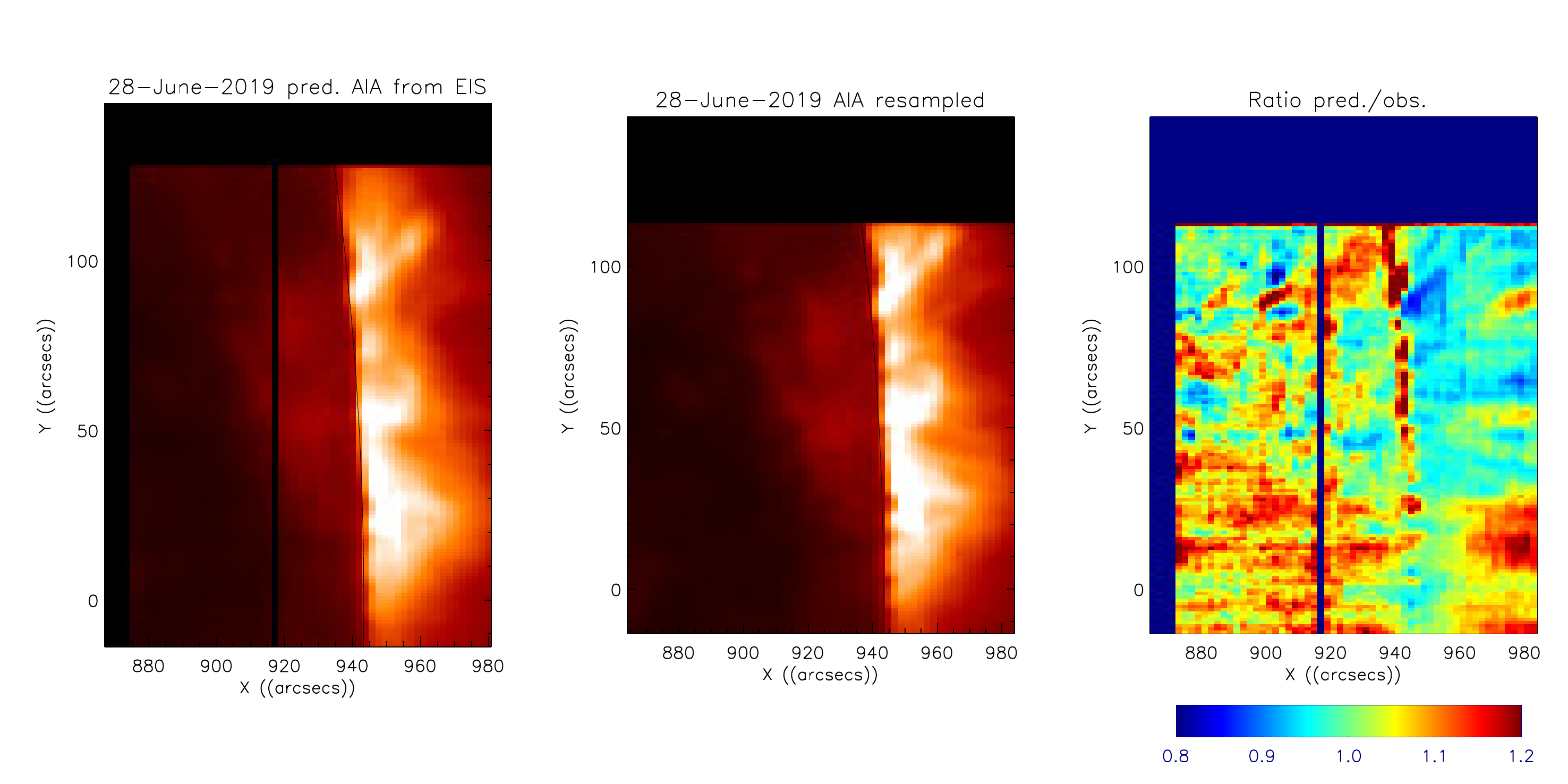}
	\caption{Results of the analysis to obtain the EIS SW radiometric calibration relative to AIA 193~\AA. }
	\label{fig:comp_eis_aia_cal}
\end{figure}

\section{EIS/AIA Co-alignment }
\label{sec:eis_aia}
  
\begin{figure}[!htbp]
	\centerline{
	\includegraphics[angle=0,width=0.5\linewidth]{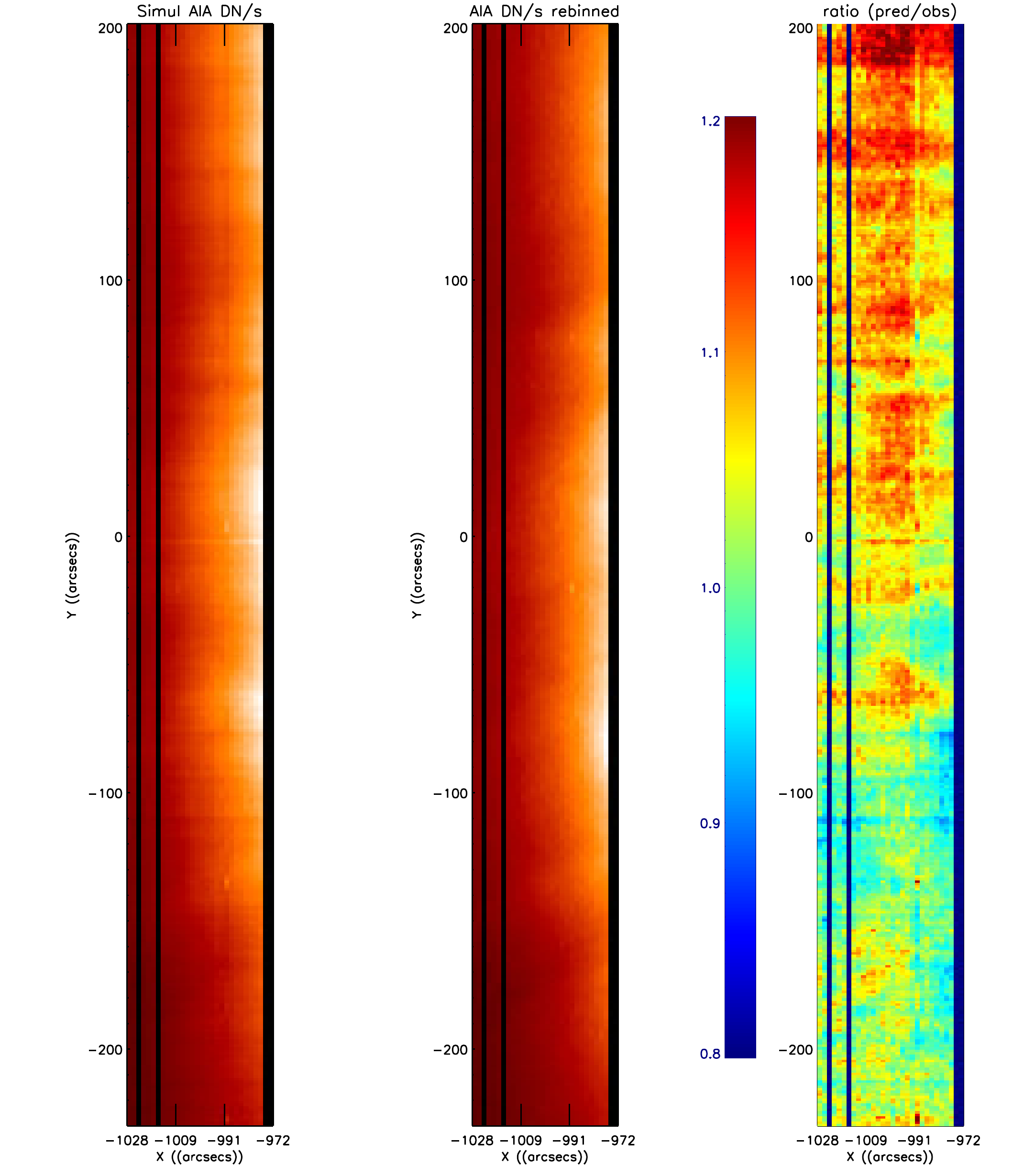}
	\includegraphics[angle=0,width=0.5\linewidth]{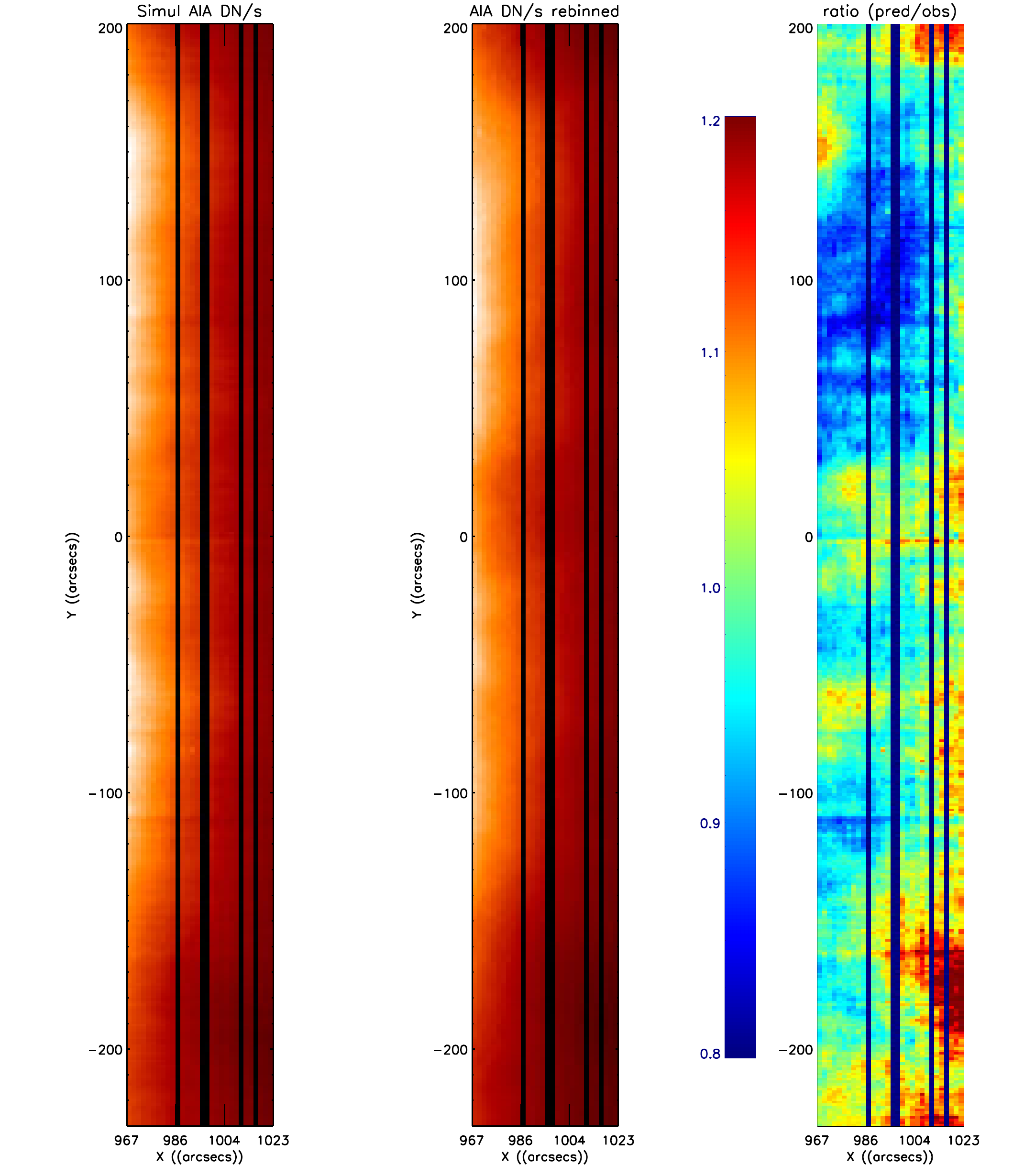}}
	\caption{Results of the analysis to obtain the EIS off-limb pointing 
	in the east and west, by comparing the predicted AIA signal with the observed one.}
	\label{fig:comp_eis_aia_fov}
\end{figure}

We took the AIA observations during the EIS rasters, and compared them
with the predicted AIA DN/s in the EIS FOV, using the updated EIS
calibration. 
We reduced the AIA count rates to the EIS spatial resolution,
and assumed a slit step size of 1.9\arcsec.
The results are shown in Figure~\ref{fig:comp_eis_aia_fov}. 
The EIS jittering, bias and point spread function limit the 
accuracy of the comparison. 
As the radial decrease in the signal is very steep,
we can accurately estimate the EIS pointing in the E-W direction
(to within 2--4\arcsec), while the N-S pointing is more uncertain
(5--10\arcsec).

\section{EIS densities on Eclipse day}

\begin{figure}[!htbp]
	\centerline{
	\includegraphics[angle=0,width=0.5\linewidth]{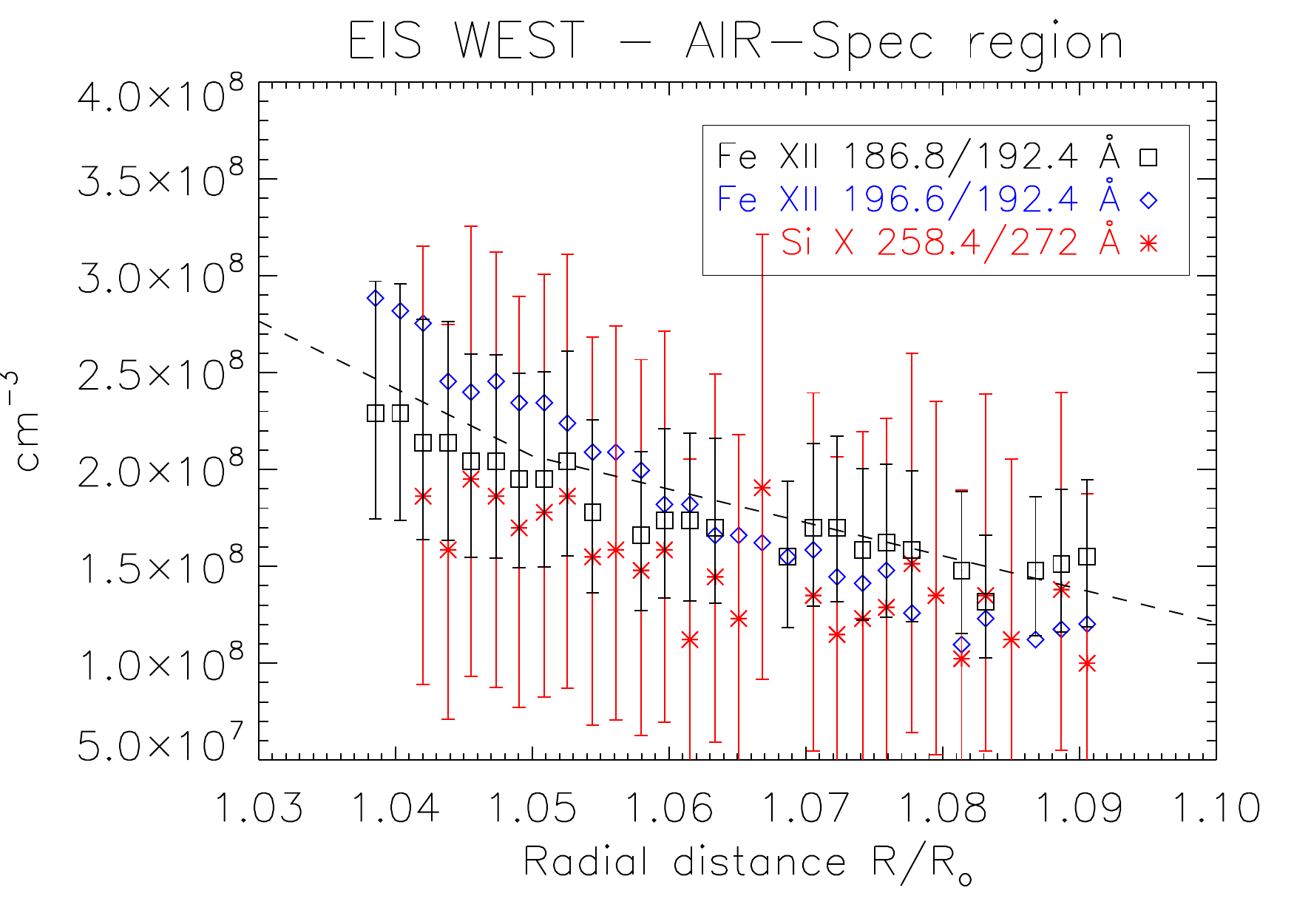}
	\includegraphics[angle=0,width=0.5\linewidth]{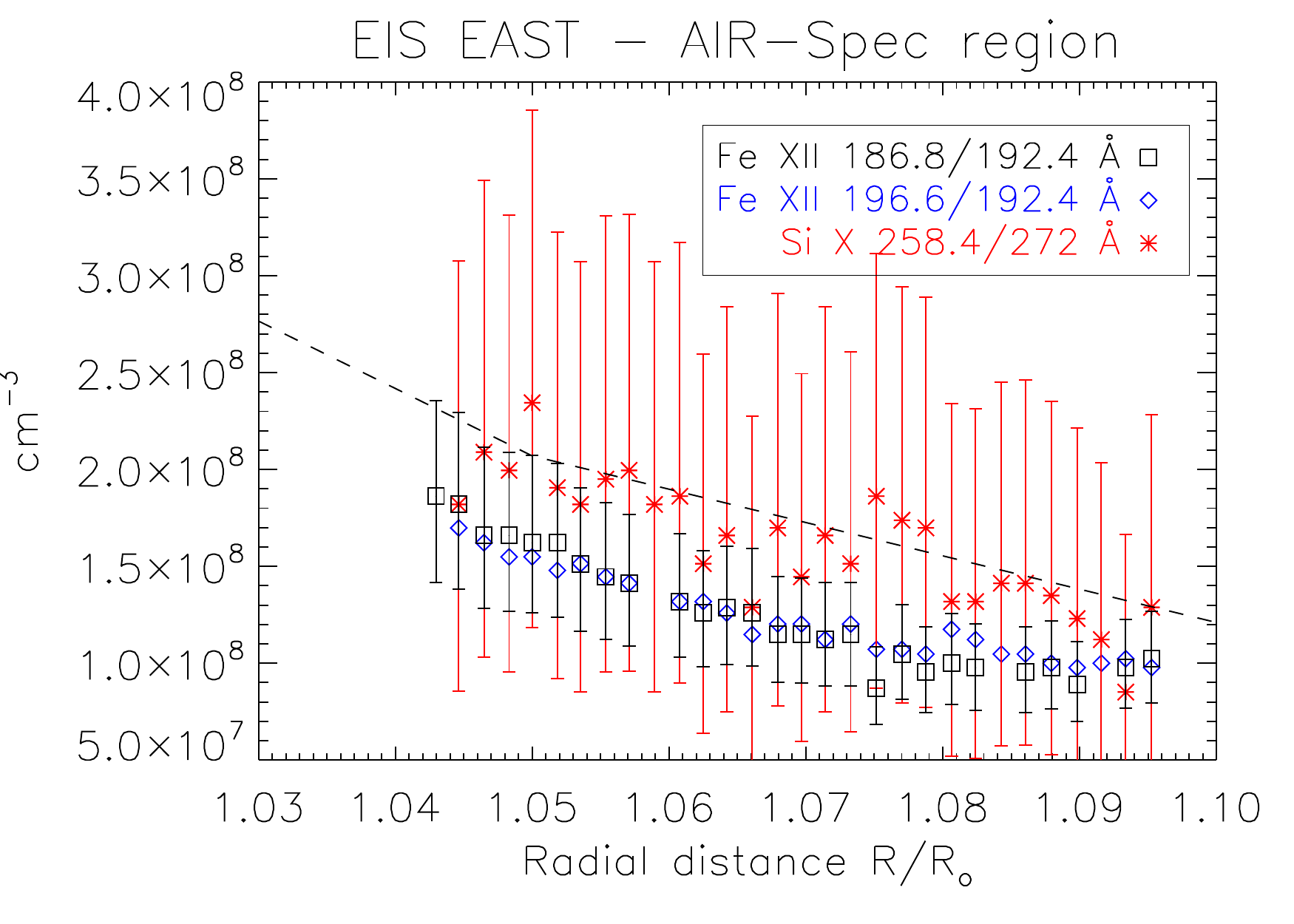}
	}
	\caption{ Electron densities obtained from a selection of EIS line ratios
	on the west and east regions. The dashed line indicates the expected radial 
	density profile for the quiet Sun, from \cite{delzanna_etal:2018_cosie}.
	}
	\label{fig:eis_ne}
\end{figure}

Figure~\ref{fig:eis_ne} shows the Electron densities obtained from a selection of EIS line ratios
on the west and east regions. The uncertainties 
are obtained by assuming a $\pm$15 \% uncertainty on the observed ratios,
projected onto the theoretical variation of the ratio, to obtain a lower and
upper limit, shown in the plot.
The 15 \%  is a reasonable estimate of the uncertainty 
due to the radiometric calibration, which is the main uncertainty. 
There is generally good agreement among the three line ratios, especially
on the west region.

\end{document}